\documentclass[12pt]{iopart}

\expandafter\let\csname equation*\endcsname\relax

\expandafter\let\csname endequation*\endcsname\relax

\usepackage{graphicx}
\usepackage{amsthm, amsmath, amssymb}
\usepackage{dsfont}
\usepackage{bm}
\usepackage{tikz}
\usepackage{tikz-3dplot}
\usepackage{tkz-euclide}
\usetikzlibrary{automata,positioning}
\usepackage{ctable}
\usepackage{array}
\newcolumntype{?}{!{\vrule width 1pt}}
\usepackage{epstopdf}
\usepackage[hyperindex,breaklinks,unicode,colorlinks=true,linkcolor=blue,citecolor=blue,urlcolor=blue]{hyperref}
\usepackage[all]{hypcap} 
\usepackage{xcolor}

\newcommand{\blue}[1]{\textcolor{black}{{#1}}}

\begin{document}

\title[Equilibrium Stochastic Delay Processes]{Equilibrium Stochastic Delay Processes}

\author{Viktor Holubec$^{1,2,\star}$, Artem Ryabov$^{1}$, Sarah A.M. Loos$^{2,3}$, Klaus Kroy$^{2}$}

\address{$^1$ Charles University,  Faculty of Mathematics and Physics, 
 Department of Macromolecular Physics, V Hole{\v s}ovi{\v c}k{\' a}ch 2, 
 CZ-180~00~Praha, Czech Republic}
\address{$^2$ Institut f\"ur Theoretische Physik, Universit\"at Leipzig,  Postfach 100 920, D-04009 Leipzig, Germany}
\address{$^3$ International Centre for Theoretical Physics, Str. Costiera 11, 34151 Trieste, Italy}
\ead{$\star$ viktor.holubec@mff.cuni.cz}

\vspace{10pt}
\begin{indented}
\item[]\today
\end{indented}

\begin{abstract}
 Stochastic processes with temporal delay play an important role in science and engineering whenever finite speeds of signal transmission and processing occur. However, an exact mathematical analysis of their dynamics and thermodynamics is available for linear models only.  We introduce a class of \textcolor{black}{stochastic delay processes with nonlinear time-local forces and linear time-delayed forces} that obey fluctuation theorems and converge to a Boltzmann equilibrium at long times. From the point of view of control theory, such ``equilibrium stochastic delay processes'' are stable and \textcolor{black}{energetically} passive, by construction. Computationally, they \textcolor{black}{provide diverse exact constraints on general nonlinear stochastic delay problems and can, in various situations,} serve as a starting point for their perturbative analysis. Physically, they admit an interpretation in terms of an underdamped Brownian particle that is either subjected to a time-local force in a non-Markovian thermal bath or to a delayed feedback force in a Markovian thermal bath. We illustrate these properties numerically for a setup  familiar from feedback cooling and point out experimental implications.
\end{abstract}

%
\vspace{2pc}

%
\submitto{\NJP}

\section{Introduction}
\label{sec:introduction}

Consider the stochastic delay differential equations (SDDEs)
\begin{eqnarray}
\dot{x}(t) &=&  v(t),
\label{eq:tde2x}\\
m\dot{v}(t) &=&  F(t) + F_{\rm D}(t-\tau) + \eta(t)
\label{eq:tde2v}
\end{eqnarray} 
with a nonlinear time-local force $F(t)=F(x,v,t)$ and a linear delay force ($\tau>0$)  
\begin{equation}
F_{\rm D}(t-\tau) = - \kappa_\tau x(t-\tau) -\gamma_\tau v(t-\tau),
    \label{eq:DelayForce}
\end{equation}
with constant coefficients $\kappa_\tau$ and $\gamma_\tau$. The dynamics is \textcolor{black}{randomly driven 
by a possibly non-Markovian, zero-mean Gaussian stochastic noise $\eta(t)$.
Intuitively, one can think of Eqs.~\eqref{eq:tde2x} and \eqref{eq:tde2v} as
describing the time evolution of the position $x(t)$ and velocity $v(t)$ of a  Brownian particle with mass $m$ and driven by the combined forces $\eta$, $F$, and $F_D$. These forces can arise from various origins, e.g., from the environment and the experimental apparatus, including some specifically tailored feedback mechanisms. Further specifications and various interpretations will be provided below.} 
Due to finite speeds of information transfer and processing and elements with slow response, such equations are ubiquitous in engineering~\cite{Kyrychko2010}, biology~\cite{Beuter1993, Yanqing1997, Novak2008} and even economics~\cite{Mackey1989, Voss2002, Stoica2005, Gao2009}. Most frequently, they are applied in modelling of feedback loops~\cite{Bechhoefer2005, Atay2010, Lakshmanan2011, Gernert2016, dissertation_of_Sarah, Baraban2013, Qian2013, Bregulla2014, Vicske2014, Mijalkov2016, Zheng2017,Zhang2017,Leyman2018, Khadka2018, Lavergne2019, Piwowarczyk2019, Bauerle2020}, neural networks~\cite{Foss1996, Marcus1989, Sompolinsky1991, Haken2007}, population dynamics~\cite{Gopalsamy2013, Mao2005}, and epidemiology~\cite{BERETTA2001,Rihan2020}.

Rising interest in SDDEs among physicists~\cite{Otto2019} is driven by recent experiments. In the so-called feedback cooling experiments with Brownian particles, one employs a feedback of the particle's past velocity to achieve a more localised state~\cite{Bushev2006,li2013millikelvin,Goldwater2019,Penny2021}. In the surging field of active matter~\cite{Ramaswamy2010, Bechinger2016, Gompper2020}, inevitable time delays in the control of robotic swarms~\cite{Mijalkov2016} led to investigations of the stability and localization of many-body systems with delayed interactions~\cite{Khadka2018, Mijalkov2016, Leyman2018, Piwowarczyk2019, Geiss2019}. In agreement with engineering practice~\cite{Kyrychko2010,Atay2010, Lakshmanan2011,Fridman2016}, it was found that delay generally introduces instabilities and oscillations into the dynamics~\cite{Khadka2018,Geiss2019} and increases stability and localization only in special cases~\cite{Bechinger2016}.

Similarly, inevitable instrumental and feedback delays in micro-manipulation experiments~\cite{Yonggun2012,Khadka2018} used to test stochastic thermodynamics~\cite{Sekimoto2010, Seifert2012} has triggered investigation of the thermodynamic aspects of SDDEs~\cite{Rosinberg2015, Rosinberg2017, Vu2019, Loos2019, dissertation_of_Sarah}. There are interesting consequences of the acausality of time-reversed processes in delay systems due to the tracking (future) history for the time-reversal. If interpreted as feedback-driven systems with information inflow, their total entropy production rate, $\dot{S}_{\rm tot}$, evaluated as a ratio of forward to backward path probabilities, is not just the sum, $\dot{S}_{\rm S} + \dot{S}_{\rm NM}$, of entropy fluxes into the system (S) and into the bath (B)~\cite{Munakata2014,Rosinberg2015,Rosinberg2017,Loos2019}. This means that the second law $\dot{S}_{\rm tot}\ge 0$ does not imply positivity of $\dot{S}_{\rm S} + \dot{S}_{\rm NM}$. These results are generic for the system~\eqref{eq:tde2x}--\eqref{eq:tde2v} with a Gaussian white noise 
$\eta(t) \propto \xi(t)$, $\left<\xi(t)\right> = 0$, $\left<\xi(t)\xi(t')\right> = \delta(t-t')$. However, explicit expressions are currently only available for linear systems \cite{Munakata2014,Rosinberg2015,Rosinberg2017,Munakata2009,Loos2019}, which fail to describe a broad range of interesting effects observed in presence of nonlinear forces~\cite{Rosinberg2015,Rosinberg2017,Loos2019}. The same can be said about the probability densities for SDDEs. They are available only for simple linear setups~\cite{Adelman1976, Fox1977, Hanggi1978, Sancho1982,  Hernandez-Machado1983, Kuchler1992, Guillouzic1999, Frank2003, McKetterick2014, Giuggioli2016,Geiss2019}, and nonlinear systems have been treated by various approximate techniques~\cite{Guillouzic1999, Frank2003, Frank2005, Mijalkov2016,Loos2017, Geiss2019, Loos2019b}. 

\textcolor{black}{Even without time delay, an exact treatment of nonlinear systems is indeed difficult. However, their stationary and relaxation properties are known exactly in thermodynamic equilibrium. In this work, we
extend this property to a certain class of SDDEs. Our results can be of interest not only to the theory of delay processes but also in applied contexts, like in control theory.}

\section{Main results}
\label{sec:results}

As our main result, we identify a class of nonlinear delay processes that admit a standard thermodynamic description, including the second law inequality $\dot{S}_{\rm tot}=\dot{S}_{\rm S} + \dot{S}_{\rm NM} \ge 0$. If not driven, they obey Boltzmann statistics in the steady state. We therefore characterize these processes as ``equilibrium delay processes''. The key idea is to accompany the time-delayed feedback force applied to the system with a suitable colored noise $\eta_\mathrm{FB}$ and interpret the resulting overall system as a particle immersed in an equilibrium reservoir and controlled by time-local external forces. Noteworthy, such feedback noise can already be realized in state-of-the-art experimental setups~\cite{DAnna2003,Murch2012,Ferialdi2019,Penny2021}. We further point out how to interpret Eqs.~\eqref{eq:tde2x} and \eqref{eq:tde2v} as a feedback-driven system and how to apply our results therein. \textcolor{black}{Altogether, we provide three complementary interpretations for the same stochastic process: a special type of system with time-delayed forces (Sec.~\ref{sec:introduction}), system with time-local forces and a heat bath with memory (Sec.~\ref{sec:mapping}), and a feedback-driven system (Sec.~\ref{sec:EFB}). They differ just in the interpretation of the individual forces on the right hand side of Eq.~\eqref{eq:tde2v}. In Tab.~\ref{tab:interpretations}, we summarize relations between the three interpretations and the definitions of the corresponding forces. In the following, we take Boltzmann's constant $k_{\rm B}$ as our unit of entropy.}

\begin{table}[]
\centering
\begin{tabular}{?c?}
\specialrule{1pt}{1pt}{1pt}
 {\bf A: Delay system} \\ 
$m\dot{v}  =  F+F_{\rm D}+\eta$\\ 
\specialrule{1pt}{1pt}{1pt}
\specialrule{1pt}{1pt}{1pt}
\emph{time-local systematic force}:
$F = F_{\rm E} + \kappa_\tau x(t) - \gamma_0 v(t)$\\
\hline
ordinary (non-feedback) external force: $F_{\rm E} = - \partial_x U(x,t) + F_{\rm N}(x,v,t)$\\
\hline
potential component of $F_{\rm E}$: $-\partial_x U(x,t)$\\
\hline
non-potential component of $F_{\rm E}$: $F_{\rm N}(x,v,t)$\\
\hline
\emph{time-delayed force}: $F_{\rm D} = - \kappa_\tau x(t-\tau) -\gamma_\tau v(t-\tau)$\\
\hline
\emph{total coloured noise from the environment and experimental apparatus}: $\eta(t)$ 
\\
\specialrule{1pt}{1pt}{1pt}
\end{tabular}

\begin{tabular}{c}
\phantom{space}
\end{tabular}

\begin{tabular}{?c?}
\specialrule{1pt}{1pt}{1pt}
 { \bf B: System with non-Markovian heat bath and time-local control }\\ 
 $m\dot{v}  = F_{\rm E} + F_{\rm F}+\eta$\\
\specialrule{1pt}{1pt}{1pt}
\specialrule{1pt}{1pt}{1pt}
\emph{time-local external force}: $F_{\rm E}$
\\
\hline
\emph{time-delayed \textcolor{black}{non-Markovian} bath friction}: \\
$F_{\rm F} = \kappa_\tau x(t) - \gamma_0 v(t) - \kappa_\tau x(t-\tau) -\gamma_\tau v(t-\tau) = \kappa_\tau x(t) - \gamma_0 v(t)  + F_{\rm D}$
\\
\hline
\textcolor{black}{total force from the non-Markovian heat bath at temperature $T$: $F_{\rm F} +\eta(t)$}
\\
\hline
\emph{coloured noise from \textcolor{black}{the non-Markovian bath}: $\eta(t)$}
\\
\hline
\textcolor{black}{heat flux into the system from the non-Markovian bath: $\dot{Q}_{\rm NM} = \left<(F_{\rm F} + \eta)\dot{x}\right>$}
\\
\specialrule{.1em}{.2em}{.2em}
\end{tabular}


\begin{tabular}{c}
\phantom{space}
\end{tabular}

\begin{tabular}{?c?}
\specialrule{1pt}{1pt}{1pt}
 {\bf C: Feedback-driven system with Markovian heat bath} \\ 
$m\dot{v}  = F_{\rm E} + F_{\rm FB} - \gamma_0 v(t)+\sqrt{2T_0\gamma_0}\xi(t)$\\
\specialrule{1pt}{1pt}{1pt}
\specialrule{1pt}{1pt}{1pt}
\emph{time-local external force}: $F_{\rm E}$\\
\hline
\emph{feedback force}:\\ \textcolor{black}{$F_{\rm FB} = \kappa_\tau x(t) - \kappa_\tau x(t-\tau) -\gamma_\tau v(t-\tau) = \kappa_\tau x(t) + F_{\rm D} + \eta_{\rm FB}(t) = F_{\rm F} + \gamma_0 v(t)$}
\\
\hline
\textcolor{black}{non-Markovian noise exerted by the feedback loop: $\eta_{\rm FB}(t) = \eta(t) - \sqrt{2T_0\gamma_0}\xi(t)$}
~~
\\
\hline
\textcolor{black}{total force from the Markovian bath at temperature $T_0$: $- \gamma_0 v(t) + \sqrt{2T_0\gamma_0}\xi(t)$}
\\
\hline
~~
\emph{time-local friction \textcolor{black}{from the Markovian bath}}: $ - \gamma_0 v(t) $
~~
\\
\hline
\emph{white noise  \textcolor{black}{from the Markovian bath}}: $\sqrt{2T_0\gamma_0}\xi(t)$ \\
\hline
\textcolor{black}{heat flux into the system from the Markovian bath: $\dot{Q}_{\rm M} = \left<(-\gamma_0 v + \eta)\dot{x}\right>$}
\\
\specialrule{1pt}{1pt}{1pt}
\end{tabular}

\caption{
Three interpretations of the delay Langevin equation~\eqref{eq:tde2v} employed in this paper and the corresponding forces. \textcolor{black}{The forces in the three interpretations yield the same change in the momentum $m\dot{v}$ and thus the same stochastic process. The term $\kappa_\tau x(t) - \gamma_0 v(t)$ in the force $F$ in the interpretation A is introduced to facilitate the reinterpretation of the delayed force $F_{\rm D}$ as part of the friction force $F_{\rm F}$ in B.} By ``ordinary external forces'' in A we mean time-local forces arising from physical interactions and thus not applied via a feedback loop.}
\label{tab:interpretations}
\end{table}

\subsection{Mapping to time-local control and non-Markovian heat bath (Tab.~\ref{tab:interpretations}B)}
\label{sec:mapping}

\textcolor{black}{In this section, we describe the reinterpretation of the delay system (Tab.~\ref{tab:interpretations}A) as an equilibrium system with memory (Tab.~\ref{tab:interpretations}B). Consider a delay system described by Eqs.~\eqref{eq:tde2x}--\eqref{eq:tde2v} with the time-local force}
\begin{equation}
F(t) = F_{\rm E}(x,v,t)+ \kappa_\tau x(t) - \gamma_0 v(t).
\label{eq:TotalForce}
\end{equation}
\textcolor{black}{
The specific form of the terms proportional to the constants $\gamma_0 > 0$ and $\kappa_\tau$ facilitates the reinterpretation of the delay force $F_D$ in Eq.~\eqref{eq:tde2v} as part of a friction force, below. The remaining force in Eq.~\eqref{eq:TotalForce},}
\begin{equation}
 F_{\rm E}(x,v,t) = - \partial_x U(x,t) + F_{\rm N}(x,v,t) 
 \label{eq:ExternalForce}
\end{equation}
is an arbitrary time-local force applied by external agents. It is composed of potential and non-potential components $- \partial_x U(x,t)$ and $F_{\rm N}(x,v,t)$. \textcolor{black}{To distinguish the force $F_{\rm E}$ from the force applied via the feedback loop in the feedback interpretation of Eqs.~\eqref{eq:tde2x}--\eqref{eq:tde2v} (Tab.~\ref{tab:interpretations}C), we call it the ``ordinary'' external force.}

Equation~\eqref{eq:tde2v} now assumes the form
\begin{equation}
m \dot{v}(t) = F_{\rm E}(t) + F_{\rm F}(t) + \eta(t)
\label{eq:vMain}
\end{equation}
with
$
F_{\rm F}(t) \equiv 
\kappa_\tau[x(t) - x(t - \tau)] - \gamma_0 v(t) - \gamma_\tau v(t-\tau).
$
It resembles the dynamical equation for the velocity of a particle subjected to an external force $F_{\rm E}$ and immersed in \textcolor{black}{a viscoelastic solvent} exerting on the particle the overall force $F_{\rm F} + \eta$ 
with systematic component (friction) $F_{\rm F}$, and stochastic component (noise) $\eta$. 

Such noise and friction can be interpreted to arise from an ordinary equilibrium heat \textcolor{black}{bath, i.e., a many-body system with infinite heat capacity in thermal equilibrium,} with a somewhat peculiar memory that gives rise to an ``echo'' in the noise and friction (Tab.~\ref{tab:interpretations}B). Notably, \textcolor{black}{for an equilibrium heat bath with a friction force linear in the variables $x$ and $v$, such as $F_{\rm F}$}, the time-reversal symmetry of the underlying microscopic dynamics implies that the friction and noise are interrelated by the so-called second fluctuation-dissipation theorem or fluctuation-dissipation relation (FDR)~\cite{Kubo1966,Felderhof1978,Zwanzig2001,Kubo2012}
\begin{equation}
 \left<\eta(t)\eta(t')\right> ={T}\, \Gamma(|t-t'|).
\label{eq:FDTGen}
\end{equation}
Here $T$ denotes the temperature and $\Gamma(t)$ is the so called friction kernel defined by the integral
\begin{equation}
  F_{\rm F}(t) = - \int_{-\infty}^t dt'\, \Gamma(t-t')v(t').
  \label{eq:friction_kernel}
\end{equation}
 For a given friction $F_{\rm F}$, the FDR~\eqref{eq:FDTGen} might imply that the noise must be complex valued. \textcolor{black}{However, in order to admit its ordinary physical interpretation and realisability in a lab,} $\eta(t)$ is required to be a real-valued function. This condition implies that its power spectrum must be non-negative,
\begin{equation}
S(\omega) = \int_{-\infty}^\infty dt 
\left<\eta(t)\eta(0)\right> \exp(-i \omega t) \ge 0.
\label{eq:PowerSpectrumGen}
\end{equation}
For the system of Eqs.~\eqref{eq:tde2x}--\eqref{eq:vMain}, the conditions~\eqref{eq:FDTGen} and \eqref{eq:PowerSpectrumGen} can be satisfied for a certain range of model parameters only, see Secs.~\ref{sec:PF} and \ref{sec:VF}. In this range, Eqs.~\eqref{eq:tde2x} and \eqref{eq:vMain} can be interpreted as describing a system with internal Hamiltonian $H = U(x,t) + m v^2/2$ acted upon by a non-potential force $F_{\rm N}$ and coupled to a non-Markovian ``equilibrium bath'' at temperature $T$. \textcolor{black}{Let us now review some general properties of this system.}

\subsection{Properties of the mapping (Tab.~\ref{tab:interpretations}B)}
\label{sec:properties}

\emph{Average thermodynamics.} If the above equilibrium mapping holds, the system's thermodynamics obeys standard relations from classical~\cite{callen1985} and stochastic~\cite{Sekimoto2010,Seifert2012} thermodynamics. Namely, the average entropy flux into the non-Markovian heat bath at temperature $T$ is given by the Claussius equality
\begin{equation}
\dot{S}_{\rm NM} = -\dot{Q}_{\rm NM}/{T}
\label{eq:SB}
\end{equation}
where $\dot{Q}_{\rm NM} = \left<(F_{\rm F} + \eta) \dot{x} \right>$ is the average heat flux from the heat bath into the system. It can also be interpreted as the work done by the bath on the system per unit time. Here and below we employ Stratonovich calculus. The averages $\left< \bullet \right>$ should be performed over many realizations of the stochastic process. 

The average heat flux is related via the first law, $d\langle H\rangle/dt = \dot{Q}_{\rm NM} + \dot{W}_{\rm E}$, to the average power input, $\dot{W}_{\rm E} = \langle \partial U/\partial t + F_{\rm N} \dot{x} \rangle$, of the system, due to external manipulations of the potential $U$ and the non-potential force $F_{\rm N}$. 
The sum of the rate of change of the system entropy, $\dot{S}_{\rm S}$, and the entropy influx $\dot{S}_{\rm NM}$ in Eq.~\eqref{eq:SB} is the total entropy production, which obeys the second law of thermodynamics~\cite{callen1985}:
\begin{equation}
\dot{S}_{\rm tot} = \dot{S}_{\rm S} + \dot{S}_{\rm NM}  \geq 0.
\label{eq:Stot}
\end{equation}

\emph{Dynamics.} Unlike a general delay system, which can exhibit over-damped, damped oscillatory, but also diverging behavior~\cite{Frank2003, McKetterick2014, Geiss2019, dissertation_of_Sarah, Atay2010, Lakshmanan2011}, systems obeying the mapping of Sec.~\ref{sec:mapping} always eventually relax into a time-independent steady state for time independent parameters, confining potential $U$, and stationary non-potential forces $F_{\rm N}$. If the latter vanishes in Eq.~\eqref{eq:ExternalForce}, the stationary probability density function (PDF) for position and velocity is given by the Gibbs canonical distribution, $p(x,v;T) = p_x(x;{T})p_v(v;{T})$, with
\begin{eqnarray}
p_x(x;{T}) &=& \exp[-U(x)/{T}]/Z_x({T}),
\label{eq:px}\\
p_v(v;T) &=& \exp(- m v^2/(2{T}))/Z_v({T}),
\label{eq:pv}
\end{eqnarray}
normalized by $Z_x({T}) = \int_{-\infty}^{\infty} dx\, \exp(-U(x)/{T})$ and $Z_v({T}) = \int_{-\infty}^{\infty} dv\, \exp(- m v^2/(2{T}))$. This is an equilibrium steady state and thus the corresponding entropy production rates $\dot{S}_{\rm S}$, $\dot{S}_{\rm NM}$, and $\dot{S}_{\rm tot}$ vanish.
For quasi-static variations of the potential, when the system PDF evolves through a set of such states, the total entropy change $\Delta S_{\rm tot} = \int_0^t \dot{S}_{\rm tot}(t') dt'$ vanishes and the entropy change in the system, $\Delta S_{\rm S}$, is exactly balanced by the entropy change in the bath, $\Delta S_{\rm NM}$. \textcolor{black}{The relaxation process to equilibrium is always accompanied by a decrease in the free energy of the system. This functional thus represents the Lyapunov function for the relaxation process that can be easily evaluated from stochastic trajectories of the system. Even stronger restrictions on the relaxation dynamics are imposed by the Evans--Searles fluctuation theorem~\cite{Evans2009,evans2016}. In contrast, besides a limited success~\cite{Maes2011}, it is currently unknown if similar general restrictions also apply to relaxation towards non-equilibrium steady states.}

The validity of these results for an arbitrary potential $U(x)$ follows from general considerations of equilibrium statistical physics~\cite{callen1985} and the FDR~\cite{Kubo1966,Felderhof1978,Zwanzig2001,Kubo2012}. However, a closed dynamical equation, e.g., of Fokker-Planck type~\cite{Risken,vanKampen}, for the PDF of a nonlinear delay process is not known~\cite{dissertation_of_Sarah} making a general direct verification difficult. In Sec.~\ref{sec:discussionA}, we provide an explicit test for the specific potential $U(x,t) = k_6 x^6/6 + k_3 x^3/3$ using Brownian dynamics (BD) simulations of Eqs.~\eqref{eq:tde2x} and \eqref{eq:tde2v}. Besides, we tested the described results for various other polynomial potentials.

We stress that the described equilibrium-like properties of equilibrium delay processes do not trivialize their dynamics. \textcolor{black}{As an example, consider a situation when the force $F$ in Eq.~\eqref{eq:tde2v} is linear in $x$ and $v$ and thus the system \eqref{eq:tde2x}-\eqref{eq:tde2v} is exactly solvable. For fixed initial conditions, one finds that the average position $\langle x(t) \rangle$ and velocity $\langle v(t) \rangle$ are identical for equilibrium ($\eta_{\rm FB}$ in Tab.~\ref{tab:interpretations} determined by the conditions~\eqref{eq:FDTGen} and \eqref{eq:PowerSpectrumGen}) 
and standard ($\eta_{\rm FB}=0$) delay processes. The four correlation functions $\langle A(t)B(0) \rangle$ for $A,B = x,v$ may then merely differ in the stationary distribution of the initial conditions.}

\emph{Fluctuation theorems.} From a stochastic-thermodynamics perspective, it is interesting to also consider a finite-speed protocol rendering the potential time-dependent. Specifically, in Sec.~\ref{sec:discussionA}, we test two fluctuation theorems for the stochastic work $w = \int_0^t \partial U(x,t')/\partial t'\, dt'$ done on the system, if $k_6 = k_6(t')$, $t' \in (0,t)$ is varied non-quasi-statically, namely the Jarzynski equality~\cite{Jarzynski1997}
\begin{equation}
\left< \exp(- w/{T})\right> = \exp(- \Delta F/{T})
\label{eq:Jarzynski}
\end{equation}
and the Crooks' fluctuation theorem~\cite{Crooks1999}
\begin{equation}
\rho_{\rm F}(w)/\rho_{\rm R}(-w) = \exp[(w - \Delta F)/{T}].
\label{eq:Crooks}
\end{equation}
Here, $\Delta F$ is the free energy difference between equilibrium states corresponding to the final and initial values of the potential, $\rho_{\rm F}$ is the probability distribution for work measured along the process when the potential changes from $U(x,0)$ to $U(x,t)$, and $\rho_{\rm R}$ is the probability distribution for work measured along the time-reversed process. For the both fluctuation theorems, the forward process departs from equilibrium. The validity of Jarzynski's equality requires the existence of initial and final Gibbs stationary states and Crooks' fluctuation theorem additionally requires the FDR and Gaussianity of the noise~\cite{Speck2007}. The described processes fulfill all these requirements and, indeed, our simulations confirm Eqs.~\eqref{eq:Jarzynski} and \eqref{eq:Crooks}.

\emph{Perturbative expansions.} Even though based on an
ad hoc choice of the noise our results represent first exact analytical solutions to \textcolor{black}{stationary PDFs for} a nonlinear SDDE. As such, they might pave the way for studying steady states and thermodynamic properties of systems controlled by more general nonlinear SDDEs. We show in Sec.~\ref{sec:response} that linear-response theory~\cite{Zwanzig2001} can be used to calculate time-dependent averages in perturbed (nonlinear) equilibrium delay systems. Besides such classical linear response, one can derive some explicit approximate formulas for specific perturbations on the level of moments calculated directly from the nonlinear system of SDDEs \eqref{eq:tde2x} and \eqref{eq:tde2v}.

\subsection{Equilibrium feedback (Tab.~\ref{tab:interpretations}C)}
\label{sec:EFB}

The formal interpretation of dynamical equations~\eqref{eq:tde2x}--\eqref{eq:tde2v} as a model for a system immersed in a non-Markovian equilibrium bath and driven by a time-local force $F_{\rm E}$, in Sec.~\ref{sec:mapping}, allowed us to utilize a wealth of known results. However, in practice, these equations usually describe feedback-driven systems in contact with a Markovian heat bath exerting a memoryless friction $-\gamma_0 v$ and Gaussian white noise $\sqrt{2 T_0\gamma_0 }\xi(t)$ 
with $\left<\xi(t)\xi(t')\right> = \delta(t-t')$. \textcolor{black}{Usually the system's environment provides such a bath.} To investigate this ``more natural'' interpretation, we rewrite the dynamical equation for the velocity as
\begin{equation}
m\dot{v}(t) =  F_{\rm E}(x,v,t)  + F_{\rm FB} - \gamma_0 v(t) + \sqrt{2T_0\gamma_0}\xi(t)
\label{eq:tde2vF}
\end{equation} 
and interpret it as describing a system immersed in a \textcolor{black}{standard, i.e., Gaussian and Markovian, heat bath at temperature $T_0$. 
This system is controlled by the time-local force $F_{\rm E}$ and the feedback force}
\begin{equation}
F_{\rm FB} = \kappa_\tau[ x(t) - x(t-\tau) ] - \gamma_\tau v(t-\tau) + \eta_{\rm FB}(t)
\label{eq:FBforce}
\end{equation}
composed of the systematic delayed component $F_{\rm F}$
and the ``feedback'' noise $\eta_{\rm FB}(t) \equiv \eta(t) - \sqrt{2T_0\gamma_0}\xi(t)$, see Tab.~\ref{tab:interpretations}C. Given that the conditions \eqref{eq:FDTGen} and \eqref{eq:PowerSpectrumGen} are fulfilled, we call this process an equilibrium feedback (EFB) process.

Importantly, the formal results concerning the system dynamics, i.e. the stationary PDFs~\eqref{eq:px} and \eqref{eq:pv}, are valid regardless of the interpretation, and thus they apply also for EFB. This means that the EFB is ideal from the point of view of passivity-based control~\cite{Ortega2013}, which is a branch of control theory that aims to balance the power delivered into the system with its dissipation. \textcolor{black}{Generic feedback can lead to divergences and instabilities when the energy influx by the feedback gradually increases the internal energy of the system. 
However, EFB processes are always stable and passive in the sense that the resulting steady states are robust against perturbations and all the energy injected into the system is dissipated.} In~\ref{appx:feedback_cooling}, we moreover show that, under realistic conditions, the temperature $T$ corresponding to the Boltzmann PDF reached by the EFB is always larger than the ambient temperature $T_0$.

The thermodynamics of EFB has to be treated with care. In particular, the total entropy production is interpretation-dependent. But the stochastic work done on the system by varying the potential remains the same, and the fluctuation theorems~\eqref{eq:Jarzynski} and \eqref{eq:Crooks} are still valid. Differences arise in the definitions of the remaining thermodynamic fluxes. With the present definition of the heat bath, the heat flux reads $\dot{Q}_{\rm M} = \langle (- \gamma_0 v + \sqrt{2T_0\gamma_0}\xi) \dot{x} \rangle$.
And, in addition to the average power $\dot{W}_{\rm E}$ delivered to the system by the potential and non-potential force\blue{s}, one has to consider also the power $\dot{W}_{\rm FB} = \langle F_{\rm FB} \dot{x} \rangle$ associated with the feedback force $F_{\rm FB}$. 

In a conventional feedback process, this power is accompanied by an information influx~\cite{Munakata2014,Rosinberg2015,Rosinberg2017,Loos2019} that, for example, allows the feedback to cool the system~\cite{Bushev2006,Goldwater2019}. \textcolor{black}{The resulting (effective) temperature of the system is then smaller than the
temperature of the ambient bath, implying a positive heat flux from the bath into the system, $\dot{Q}_{\rm M} > 0$. In a steady state, the conventional feedback is thus able to cool the ambient bath by extracting the power $-\dot{W}_{\rm FB} = \dot{Q}_{\rm M} > 0$ from it. However,} for an arbitrary force $F_{\rm E}$, the second law~\eqref{eq:Stot} together with the relation $\dot{Q}_{\rm NM} = \dot{Q}_{\rm M} + \dot{W}_{\rm FB} = - T \dot{S}_{\rm NM}$ imposes an upper bound $\dot{Q}_{\rm M} \le T \dot{S}_{\rm S} - \dot{W}_{\rm FB}$ on the heat delivered from the bath to the system via the EFB. And, in~\ref{appx:feedback_cooling}, we show that under equilibrium conditions, $\partial U/\partial t = F_{\rm N} = \dot{S}_{\rm S} =0$, \textcolor{black}{the EFB brings the system to an effective temperature, $T$, larger than the ambient temperature, $T_0$. Hence, the heat flux $\dot{Q}_{\rm M}$ is always negative, the EFB performs net work on the system, $\dot{W}_{\rm FB} = - \dot{Q}_{\rm M} > 0$, and it eventually heats the ambient bath.} 
This means that the EFB cannot be used for standard (zero non-potential force and time-independent potential) feedback cooling of the system~\cite{Bushev2006,Goldwater2019}.

Sections~\ref{sec:PF} and~\ref{sec:VF} clarify when EFB can be realized with time-delayed forces depending on either the earlier position or velocity\textcolor{black}{, i.e., when the corresponding feedback noise $\eta_{\rm FB}$ in Tab.~\ref{tab:interpretations}C can be constrained to be real valued}. The technical details are given in~\ref{appx:noises_in_reality}.
The resulting parameter regimes where the EFB can be realized in these two situations are depicted in phase diagrams (Figs.~\ref{fig:PDx} and \ref{fig:PD}). Equilibrium feedback with time-delayed forces depending on both delayed position and velocity can be investigated in a similar manner, but the corresponding phase diagram becomes three-dimensional. In Sec.~\ref{sec:discussionA}, we verify the validity of our theoretical results by a Brownian dynamics (BD) simulation of the equilibrium velocity feedback. In Sec.~\ref{sec:response}, we discuss several perturbative expansions pushing the theory beyond the parameter regime of the equilibrium delay processes. We conclude in Sec.~\ref{sec:conclusion}.

\section{Equilibrium position feedback}
\label{sec:PF}

Let us now consider the situation of the position-dependent feedback force ($\gamma_\tau = 0$ in Eq.~\eqref{eq:FBforce}) 
\begin{equation}
F_{\rm FB} = \kappa_\tau[ x(t) - x(t-\tau) ] + \eta_{\rm FB}(t) = F_{\rm F} + \gamma_0 v(t) + \eta_{\rm FB}(t).
\label{eq:FBforceX}
\end{equation}
The generalized friction force $
F_{\rm F} = 
\kappa_\tau[x(t) - x(t - \tau)] - \gamma_0 v(t)
$ 
can be written using the friction kernel 
\begin{equation}
\Gamma(t) = \left[2\gamma_0 \delta(t) - \kappa_\tau \Theta(\tau - t)\right]\Theta(t),
\label{eq:Gammaxv}
\end{equation}
where $\Theta(.)$ denotes the Heaviside step function.
This result can be verified by direct substitution into Eq.~\eqref{eq:friction_kernel} and integrating the term including velocity $v(t) = \dot{x}(t)$ by parts, cf. Eq.~(9.14) in~Ref.~\cite{dissertation_of_Sarah}.

The conditions~\eqref{eq:FDTGen} and \eqref{eq:PowerSpectrumGen} on the EFB imply that the friction kernel~\eqref{eq:Gammaxv} and the total noise $\eta(t) = \eta_{\rm FB}(t)+\sqrt{2T_0\gamma_0}\xi(t)$ (see Tab.~\ref{tab:interpretations}A) must obey the FDR,
\begin{equation}
  \left<\eta(t)\eta(t')\right>/{T} = 2 \gamma_0\delta(t-t')
- {\kappa}_\tau \Theta(\tau - |t-t'|),
\label{eq:noiseTCFxv}
\end{equation}
and that the corresponding power spectrum must be non-negative,
\begin{equation}
S(\omega) = 2\left[{\gamma}_0 - {\kappa}_\tau\tau \frac{\sin (\omega\tau)}{\omega\tau}\right] \ge 0.
\label{eq:PSx}
\end{equation}
Using $\max[\sin(x)/x] = 1$ and $\min[\sin(x)/x] \approx -0.22$, this implies the inequalities
\begin{equation}
0 \le \max({\kappa}_\tau, -0.22 {\kappa}_\tau) \tau \le {\gamma}_0,
\label{eq:inequality_x}
\end{equation}
which specify the parameter regime where the noise $\eta(t)$ satisfying the FDR~\eqref{eq:noiseTCFxv} can actually be realized in the lab (for details of noise realization, see~\ref{appx:noises_in_reality}).
The inequalities require non-negative $\gamma_0$ which is always fulfilled in the EFB interpretation, where $\gamma_0$ measures the strength of the background friction. For $\gamma_0\ge0$, the inequalities~\eqref{eq:inequality_x} bounds the feedback strength ${\kappa}_\tau$ as $-\gamma_0/0.22 \le {\kappa}_\tau \le \gamma_0$. 

\begin{figure}[t!]	
\centering
\begin{tikzpicture}
	\node (img1)  {\includegraphics[width=0.6
	\columnwidth]{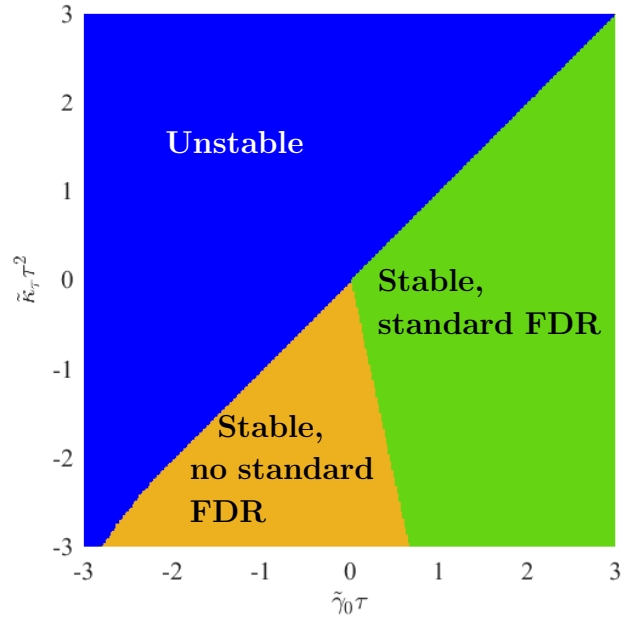}};
	\node[above=of img1, node distance=0.0cm, yshift=-6.2cm,xshift=2.2cm,text width=3cm] {\bf Stable,\\ standard FDR};
	\node[above=of img1, node distance=0.0cm, yshift=-8.7cm,xshift=-0.3cm,text width=3cm] {\bf \phantom{+}Stable,\\ no standard FDR};
	\node[above=of img1, node distance=0.0cm, yshift=-3.8cm,xshift=-1.2cm, white] {\bf Unstable};
\end{tikzpicture}
	\caption{Phase diagram of the position feedback  in the reduced variables $\tilde{\kappa}_\tau = \kappa_\tau/m$ and $\tilde{\gamma}_0 = \gamma_0/m$. In the FDR region, $\tilde{\gamma}_0 \ge \max(\tilde{\kappa}_\tau, -0.22 \tilde{\kappa}_\tau) \tau$ and the system has a positive relaxation time $t_R$. Then the system is stable for arbitrary delay and it is possible to drive it by an equilibrium position feedback. In the no-FDR region, $\tilde{\gamma}_0 < \max(\tilde{\kappa}_\tau, -0.22 \tilde{\kappa}_\tau) \tau$ and $t_{\rm R} > 0$, the system reaches a stable steady state, but the equilibrium position feedback cannot be realized in practice. In the unstable region ($t_{\rm R} < 0$), the velocity exhibits exponentially diverging oscillations \textcolor{black}{due to large time delays and thus no steady state exists. For $\tau=0$, the process is stable.}}
	\label{fig:PDx}	
\end{figure}

Under these conditions, the equilibrium position feedback fulfills all the properties described in Sec.~\ref{sec:results}. \textcolor{black}{In particular it eventually yields the stable equilibrium distribution~\eqref{eq:px} and \eqref{eq:pv} whenever $\partial U/\partial t = F_{\rm N} = 0$. However, the time delay in a general feedback may yield diverging trajectories for certain parameter values. As an independent check that the  parameter regime~\eqref{eq:inequality_x} allowing for equilibrium position feedback always leads to stable stationary solutions, we investigate the overall stability of position feedback described by Eqs.~\eqref{eq:tde2vF} and \eqref{eq:FBforceX} for the case $F_{\rm E} = 0$, which can be inspected analytically.}

Specifically, the process~\eqref{eq:tde2vF} eventually reaches a stable steady state if all the corresponding relaxation times, $t_{\rm R}$, are positive. To calculate them, we substitute the feedback force~\eqref{eq:FBforceX}, $F_{\rm E} = 0$ , and $v = \dot{x}$ in Eq.~\eqref{eq:tde2vF}, set $\eta(t) = 0$, and solve the resulting equation using the exponential ansatz $x = \exp(-\lambda t/\tau)$~\footnote{One can analogously treat systems with a potential $U$, by linearising it around a (local) minimum and absorbing the resulting linear time-local force into $\kappa_\tau$.}. The obtained transcendental equation
\begin{equation}
m\lambda^2 = \gamma_0\tau \lambda
- \kappa_\tau \tau^2 \left[\exp(\lambda)
- 1 \right]
\label{eq:lambda}
\end{equation}
can in general only be solved numerically and has infinitely many solutions. As the relaxation time of the system, $t_{\rm R}$, we identify the smallest $\tau/\Re[ \lambda]$ solving Eq.~\eqref{eq:lambda}, where $\Re[\bullet]$ denotes the real part. The system eventually relaxes into a stable steady state with $\langle v(t) \rangle=0$ if $t_{\rm R} > 0$. An approximate explicit solution to Eq.~\eqref{eq:lambda} can be obtained in the limit of small delay. Expanding the friction $F_{\rm F}$ in Eq.~\eqref{eq:FBforceX} up to the first order in $\tau$, we get
\begin{equation}
m\dot{v}(t) \approx - (\gamma_0 - \kappa_\tau \tau) v(t) + \sqrt{2T_0\gamma_0}\xi(t).
\label{eq:approx_stable_v}
\end{equation}
The last two terms can be interpreted as a noise and friction from an equilibrium bath with friction coefficient $\gamma_0 - \kappa_\tau \tau$, \textcolor{black}{which yields stable dynamics where all the energy injected into the system by the feedback is dissipated (passive dynamics)  if}
\begin{equation}\label{eq:approx_stable_v_cond}
 \kappa_\tau \tau \leq \gamma_0.
\end{equation}
The conditions for the general case are depicted in Fig.~\ref{fig:PDx}. Indeed, the
whole parameter regime where the EFB can be defined according to the FDR~\eqref{eq:noiseTCFxv} (green) is found to be stable. This shows that, as expected, EFB is passive and stable. 
Nevertheless, the regime of stability, $t_{\rm R}>0$, is broader (orange). Noteworthy, the system can be stable even for $\gamma_0<0$ if the feedback strength $\kappa_\tau$ is also sufficiently negative. This could have been anticipated from the approximate condition for stability~\eqref{eq:approx_stable_v_cond}, which predicts the boundary between stable and unstable regimes for $\tilde{\gamma}_0\tau \gtrapprox -2$ remarkably well. The approximate dynamics allows to define an effective FDR with an effective temperature $T_{\rm eff} = T_0/(1-\kappa_\tau \tau/\gamma_0)$. However, beyond the small delay approximation, the existence of such effective FDR is not guaranteed. For smaller values of $\tilde{\gamma}_0$, higher order terms in the delay make the system more unstable than expected from the linear analysis.  In the unstable regime (blue), the mean velocity exhibits exponentially increasing oscillations~\cite{McKetterick2014,Geiss2019}.

\section{Equilibrium velocity feedback}
\label{sec:VF}

Next, we perform the same analysis as in the previous section for the velocity-dependent feedback force ($\kappa_\tau = 0$ in Eq.~\eqref{eq:FBforce}) 
\begin{equation}
F_{\rm FB} = - \gamma_\tau v(t-\tau) + \eta_{\rm FB} = F_{\rm F} + \gamma_0 v(t) + \eta_{\rm FB}.
\label{eq:FBforceV}
\end{equation}
The friction $F_{\rm F} = -\gamma_\tau v(t-\tau) - \gamma_0 v(t)$ now corresponds to the friction kernel 
\begin{equation}
\Gamma(t) = \left[2\gamma_0 \delta(t) + \gamma_\tau \delta(t-\tau)\right]\Theta(t).
\label{eq:FK}
\end{equation}
in Eq.~\eqref{eq:friction_kernel}. The FDR relation~\eqref{eq:FDTGen} for the total noise $\eta(t)= \eta_{\rm FB}(t)+\sqrt{2T_0\gamma_0}\xi(t)$ (see Tab.~\ref{tab:interpretations}A) now reads
\begin{equation}
 \left<\eta(t)\eta(t')\right>/{T} = 2\gamma_0\delta(t-t') + \gamma_\tau \delta(|t-t'|-\tau),
\label{eq:noiseTCFv}
\end{equation}
and thus the condition following from the positivity of the power spectrum \eqref{eq:PowerSpectrumGen} reads
\begin{equation}
S(\omega) = 2\left[\gamma_0 + \gamma_\tau \cos(\omega\tau)\right] \ge 0.
\label{eq:PSv}
\end{equation}
The equilibrium velocity feedback thus can be realized if the inequality
\begin{equation}
0 \le |\gamma_\tau| \le \gamma_0,     
\label{eq:inequalities_gamma}
\end{equation} 
holds (for detail of the realization, see~\ref{appx:noises_in_reality}). As we have seen for the equilibrium position feedback, $\gamma_0$  must be non-negative, which is fulfilled in the EFB interpretation. For ${\gamma}_0 \ge 0$, the inequalities~\eqref{eq:inequalities_gamma} impose that the amplitude $\gamma_\tau$ of the delayed component of the friction can not exceed that of the Markov component. Different from the corresponding inequality for $\kappa_\tau$ in the position feedback, this condition is now symmetric with respect to $\gamma_\tau = 0$. 

Similarly as in the case of the position feedback, we inspect the region of stability of the general linear velocity feedback for $F_{\rm E} = 0$ and compare it to the region~\eqref{eq:inequalities_gamma} allowing to realize the stable equilibrium feedback. To this end, we insert the feedback force \eqref{eq:FBforceV} and $F_{\rm E} = 0$ in Eq.~\eqref{eq:tde2vF}, set $\eta(t) = 0$, and solve the resulting equation using the exponential ansatz $v(t) = \exp(-t/t_{\rm R} + i\omega t)$, with real parameters $t_{\rm R}$ and $\omega$. Solving the resulting algebraic equation for the relaxation time $t_{\rm R}$, we find
\begin{equation}
t_{\rm R} = \frac{\tau}{\Re\left(\tilde{\gamma}_0 \tau - {\rm W}\left[-\tilde{\gamma}_\tau \tau \exp(\tilde{\gamma}_0 \tau)\right]\right)},
\label{eq:relax_time0}
\end{equation}
where ${\rm W}[.]$ stands for the Lambert ${\rm W}$ function, $\Re(.)$ denotes the real part, and $\tilde{\gamma}_0 = \gamma_0/m$ and $\tilde{\gamma}_\tau = \gamma_\tau/m$. The Lambert ${\rm W}$ function is a multivalued function and, in order to assess stability of the system, we numerically determine the smallest $t_{\rm R}$ resulting from Eq.~\eqref{eq:relax_time0}. In this case, the small-delay expansion of the friction $F_{\rm F}$ in Eq.~\eqref{eq:FBforceV} yields
\begin{equation}
m\dot{v}(t) \approx - (\gamma_0 + \gamma_\tau) v(t) + \gamma_\tau \tau \dot{v}(t) + \sqrt{2T_0\gamma_0}\,\xi(t)
\label{eq:smallTauv}
\end{equation}
and thus it suggest that the system will be stable for $\gamma_\tau > - \gamma_0$ (positive effective friction coefficient) and $\gamma_\tau \tau/m < 1$ (positive effective mass). It also allows to define an effective FDR with an effective temperature $T_{\rm eff} = T_0/(1+\gamma_\tau/\gamma_0)$ valid for small delays.

\begin{figure}[t!]	
\centering
\begin{tikzpicture}
	\node (img1)  {\includegraphics[width=0.6\columnwidth]{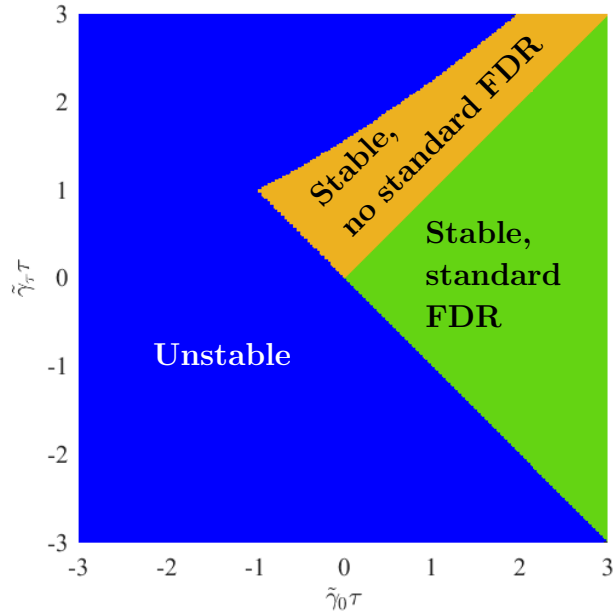}};
	\node[above=of img1, node distance=0.0cm, yshift=-6.2cm,xshift=2.9cm,text width=3cm] {\bf Stable,\\ standard\\ FDR};
	\node[above=of img1, node distance=0.0cm, yshift=-3.2cm,xshift=2.3cm, rotate=45,text width=5cm] {\bf Stable,\\ no standard FDR};
	\node[above=of img1, node distance=0.0cm, yshift=-6.7cm,xshift=-1.3cm,white] {\bf Unstable};
\end{tikzpicture}
	\caption{Phase diagram of the velocity feedback in the reduced variables $\tilde{\gamma}_0 =  \gamma_0/m$ and $\tilde{\gamma}_\tau = \gamma_\tau/m$. In the FDR region, $0 \le |\gamma_\tau| \le \gamma_0$ and the system has a positive relaxation time $t_R$. Then the system is stable for arbitrary delay and it is possible to drive it by an equilibrium velocity feedback (EFB). In the no-FDR region, $\gamma_0 \le |\gamma_\tau|$ and $t_{\rm R} > 0$, the system reaches a stable steady state but the EFB cannot be realized in practice. In the unstable region ($t_{\rm R} < 0$), the velocity exhibits exponentially diverging oscillations \textcolor{black}{due to large time delays} and thus no steady state exists.}
	\label{fig:PD}	
\end{figure}

These formulas correctly yield the bottom boundary between the unstable and stable regions in the phase diagram generated using the exact conditions \eqref{eq:inequalities_gamma} and \eqref{eq:relax_time0} depicted in Fig.~\ref{fig:PD}. As for the position feedback, the region where the FDR~\eqref{eq:noiseTCFxv} and thus the equilibrium velocity feedback can be defined (green) is indeed stable. And the regime of stability, $t_{\rm R}>0$, is broader than the FDR regime and still extends to region of negative friction coefficients $\gamma_0<0$ (orange). In the unstable regime (blue), the mean velocity again exhibits exponentially increasing oscillations~\cite{McKetterick2014,Geiss2019}.

\section{Demonstration of equilibrium velocity feedback}
\label{sec:discussionA}

Let us now discuss a specific realization of the equilibrium velocity feedback and show that it indeed has all the properties described in Sec.~\ref{sec:results}. As detailed in~\ref{appx:noises_in_reality} a possible (parsimonious) form of the total noise $\eta(t) = \eta_{\rm FB}(t)+\sqrt{2T_0\gamma_0}\xi(t)$, which fulfills the FDR~\eqref{eq:noiseTCFv} for equilibrium velocity feedback, is obtained by setting $\eta_{\rm FB}(t) = \sqrt{\alpha_\tau} \xi (t-\tau)$. The parameters of the corresponding feedback force $F_{\rm FB}(t)$ \eqref{eq:FBforceV},
\begin{equation}
F_{\rm FB}(t) = -\gamma_\tau v(t-\tau) + \sqrt{\alpha_\tau} \xi(t-\tau),
\label{eq:GF}
\end{equation}
can be tuned to represent various equilibrium velocity delay process. As a benchmark for the EFB, we consider three processes distinguished by values of the coefficients $\gamma_\tau$ and $\alpha_\tau$ above: (i) equilibrium process (EQ) with $\gamma_\tau = \alpha_\tau = 0$ and thus $F_{\rm FB}(t) = 0$; (ii) non-equilibrium (generic) velocity feedback (NEFB) with $\gamma_\tau > 0$ and $\alpha_\tau = 0$ and thus $F_{\rm FB}(t) = -\gamma_\tau v(t-\tau)$; and (iii) equilibrium velocity feedback (EFB) with $\gamma_0 \ge \gamma_\tau > 0$ and $\alpha_\tau > 0$ obeying Eq.~\eqref{eq:alpha0}. The last condition is compatible with an equilibrium state at arbitrary temperature $T>T_0$, if we set
\begin{eqnarray}
\alpha_\tau &=& 2\gamma_0 \left(T-T_0\right),
\label{eq:alphatau}\\
\gamma_\tau &=& \pm 2\gamma_0\sqrt{\frac{ T_0}{T}}\sqrt{1 - \frac{T_0}{T} } ,
\end{eqnarray}
where $T_0/T \le 1$. Thus, in agreement with the discussion in~\ref{appx:feedback_cooling}, the additional noise present in the EFB always agitates or ``heats'' the system above the ambient temperature $T_0$. 
Note that the above expressions do not depend on the delay $\tau$.

\subsection{Dynamics}
\label{sec:performanceEVF}

To gain intuition concerning the behavior of the equilibrium velocity feedback process, we
now consider the specific system obeying Eqs.~\eqref{eq:tde2x} and \eqref{eq:tde2vF} with the feedback force~\eqref{eq:GF} and the force $F_{\rm E} = -\partial U/\partial x$ induced by the potential
\begin{equation}
U(x) = \frac{k_6}{6}x^6 + \frac{k_3}{3}x^3.
\label{eq:Unonl}
\end{equation}
We solve the dynamical equations using BD simulations for the NEFB, EFB and EQ described above. In all our illustrations, we use $1/\tilde{\gamma}_0$ as our time unit and $\sqrt{T_0/m}$ as our length unit. Velocity is thus measured in units of $\tilde{\gamma}_0\sqrt{T_0/m}$. We show results from BD simulations for the two parameter sets $(\tilde{\gamma}_0 \tau,\tilde{\gamma}_\tau \tau) \approx (0.28,0.28)$ and $(\tilde{\gamma}_0 \tau,\tilde{\gamma}_\tau \tau) \approx (0.42,0.24)$. The first one yields fast relaxation of both feedback processes for $U = 0$. The second one is optimised to provide small velocity variance for EFB for $U=0$. For more details, see~\ref{appx:parameterChoice}.

\begin{figure}[t!]	
\centering
\begin{tikzpicture}
	\node (img1)  {\includegraphics[width=0.52\columnwidth]{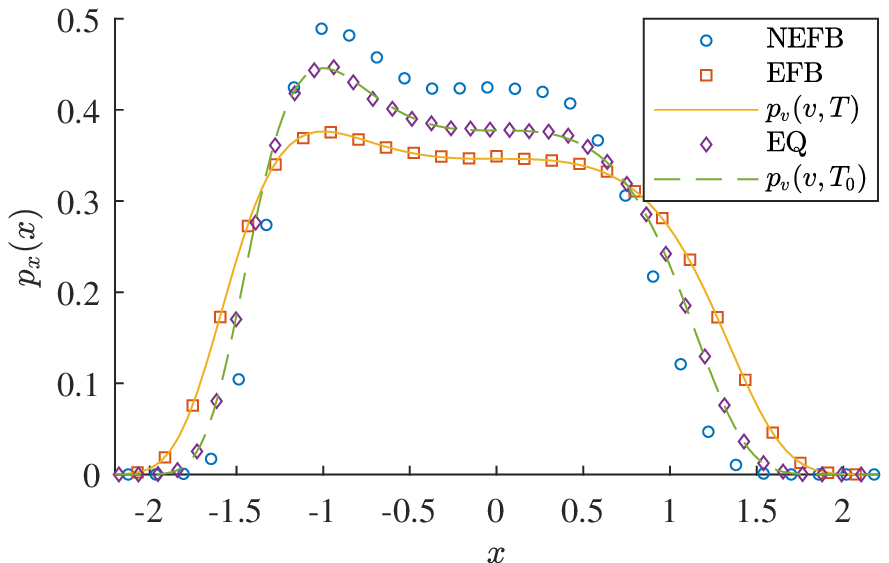}};
	\node[right=of img1, node distance=0.0cm, yshift=0cm,xshift=-1.6cm] (img2)
	{\includegraphics[width=0.52\columnwidth]{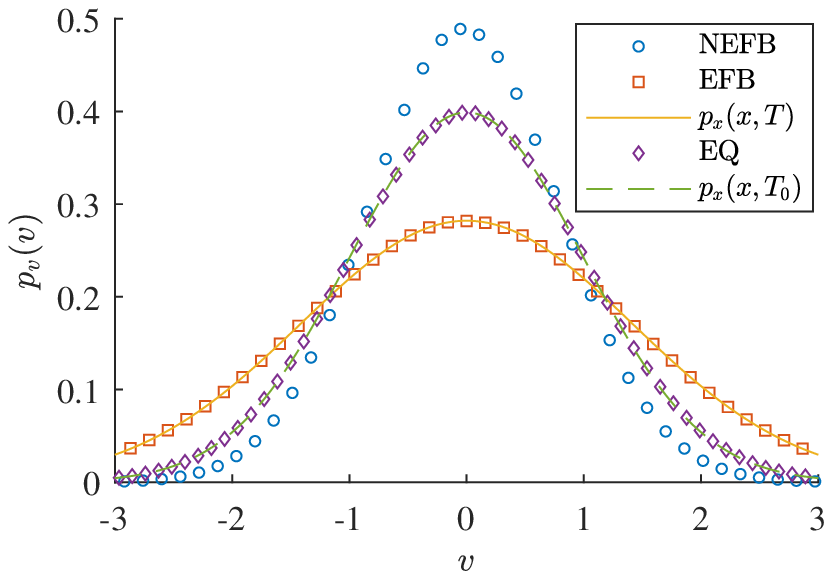}};
	\node[above=of img1, node distance=0.0cm, yshift=-1.9cm,xshift=-2.2cm] {(a)};
	\node[above=of img2, node distance=0.0cm, yshift=-1.9cm,xshift=-2.2cm] {(b)};	
\end{tikzpicture}
	\caption{Stationary PDFs for $x$ (a) and $v$ (b) for the potential~\eqref{eq:Unonl} with $k_6=k_3=1$
	in the parameter regime $\tau \approx 0.28$, $\tilde{\gamma}_\tau\tau \approx 0.28$ (optimized for relaxation times of feedback processes with $U=0$, see~\ref{appx:parameterChoice}). The data for the equilibrium process (EQ) and equilibrium velocity feedback (EFB) perfectly agree with the corresponding Boltzmann PDFs~\eqref{eq:px} and \eqref{eq:pv}.
Concerning the PDFs for generic velocity feedback (NEFB), no exact analytical formula for the shown PDFs is known. For all figures, simulation data was obtained from 50000 trajectories of length 10 with time-step 0.001.}
	\label{fig:PdfT}	
\end{figure}

\begin{figure}[t!]	
\begin{tikzpicture}
	\node (img1)  {\includegraphics[width=0.52\columnwidth]{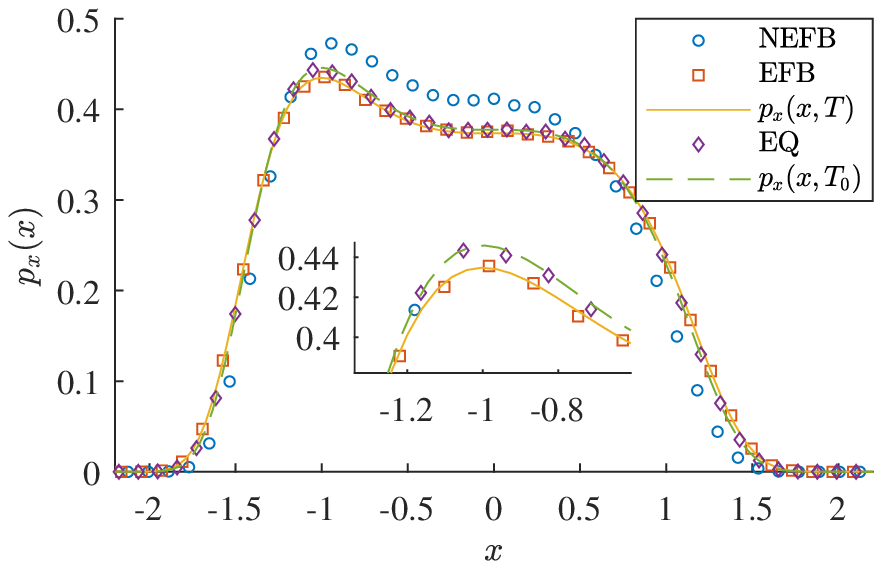}};
	\node[right=of img1, node distance=0.0cm, yshift=0cm,xshift=-1.6cm] (img2)
	{\includegraphics[width=0.52\columnwidth]{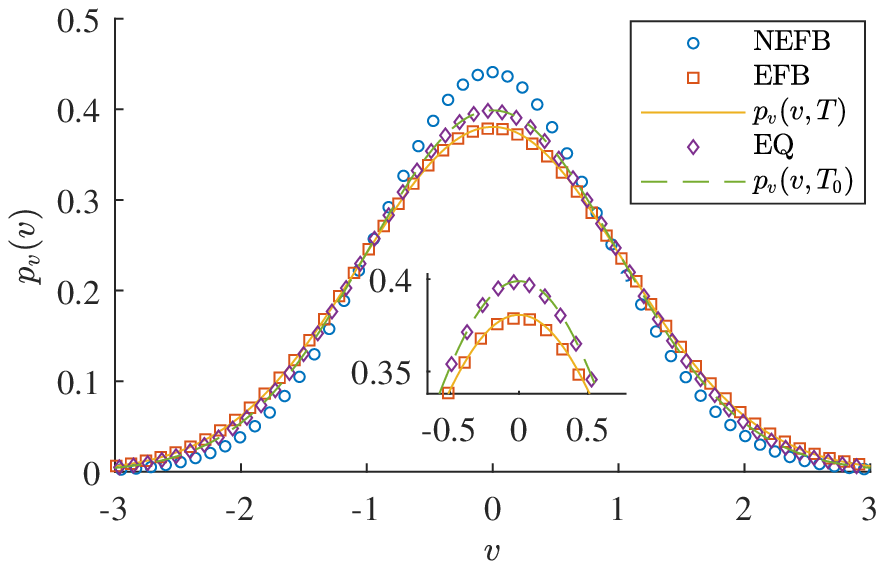}};
	\node[above=of img1, node distance=0.0cm, yshift=-1.9cm,xshift=-2.2cm] {(a)};
	\node[above=of img2, node distance=0.0cm, yshift=-1.9cm,xshift=-2.2cm] {(b)};	
\end{tikzpicture}
	\caption{The same as in Fig.~\ref{fig:PdfT} 
	in the parameter regime $\tau \approx 0.42$, $\tilde{\gamma}_\tau \tau \approx 0.24$ (optimized for velocity variance of EFB, see~\ref{appx:parameterChoice}). Other parameters are the same as in Fig.~\ref{fig:PdfT}. The insets magnify the regions around the global maxima of the PDFs. The data for the equilibrium process (EQ) and equilibrium velocity feedback (EFB) again perfectly agree with the corresponding Boltzmann PDFs~\eqref{eq:px} and \eqref{eq:pv}.
}
	\label{fig:PdfTVar}	
\end{figure}
    
\begin{figure}[t!]	
\begin{tikzpicture}
	\node (img1)  {\includegraphics[width=1.0\columnwidth]{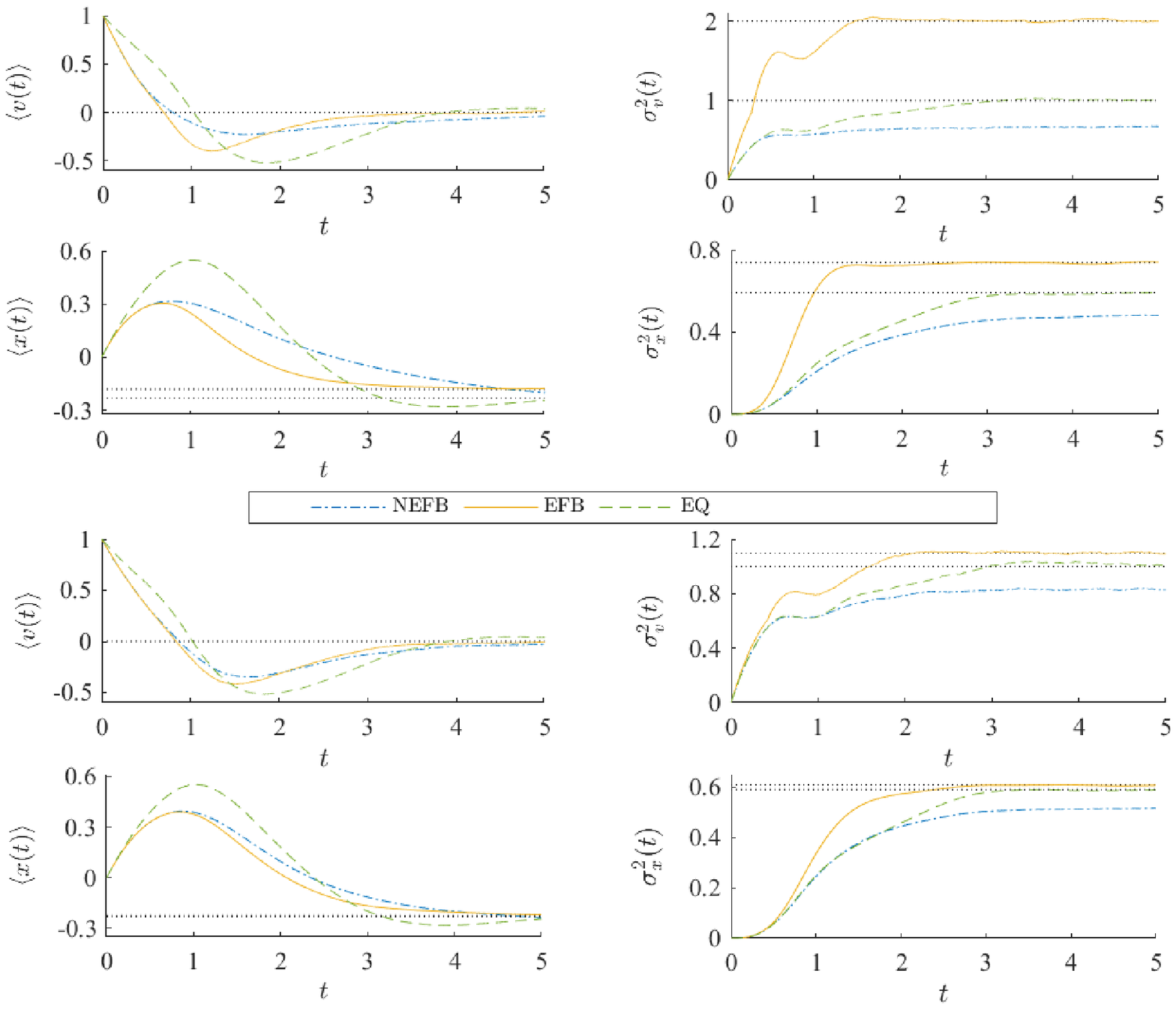}};
	\node[above=of img1, node distance=0.0cm, yshift=-1.9cm,xshift=-0.7cm] {(a)};
	\node[above=of img1, node distance=0.0cm, yshift=-3.35cm,xshift=6.8cm] {(b)};	
		\node[above=of img1, node distance=0.0cm, yshift=-4.65cm,xshift=-0.7cm] {(c)};
	\node[above=of img1, node distance=0.0cm, yshift=-6.15cm,xshift=6.8cm] {(d)};	
		\node[above=of img1, node distance=0.0cm, yshift=-8.20cm,xshift=-0.7cm] {(e)};
	\node[above=of img1, node distance=0.0cm, yshift=-9.65cm,xshift=6.8cm] {(f)};	
		\node[above=of img1, node distance=0.0cm, yshift=-10.90cm,xshift=-0.7cm] {(g)};
	\node[above=of img1, node distance=0.0cm, yshift=-12.55cm,xshift=6.8cm] {(h)};	
\end{tikzpicture}
	\caption{Relaxation dynamics of first two central moments of velocity and position for NEFB, EFB and EQ processes departing (with certainty) from the initial condition $x=0$, $v=1$ for $t \le 0$. In (a-d) we show results for parameters used in Fig.~\eqref{fig:PdfT} and in (e-h) for those used in Fig.~\eqref{fig:PdfTVar}. The horizontal lines depict stationary values of the shown moments obtained analytically using the Boltzmann distributions \eqref{eq:px} and \eqref{eq:pv} in Figs.~\ref{fig:PdfT} and \ref{fig:PdfTVar}.} 
	\label{fig:NonlinearCase}	
\end{figure}

 In Figs.~\ref{fig:PdfT} and \ref{fig:PdfTVar} we show the stationary PDFs for $x$ and $v$ obtained for the first and second parameter set, respectively. In both figures, the simulated PDFs for EQ and EFB perfectly overlap with the corresponding analytical Boltzmann PDFs~\eqref{eq:px} and \eqref{eq:pv} providing numerical evidence for our claims in Sec.~\ref{sec:results}. As expected, the position and velocity fluctuations are always smallest for the NEFB and largest for the EFB. 

To compare the relaxation dynamics of the three processes, we show in Fig.~\ref{fig:NonlinearCase} the corresponding mean values $\left<x\right>$ and $\left<v\right>$
and variances $\sigma_x^2$ and $\sigma_v^2$ as functions of time for the initial condition $(v,x)=(1,0)$ for $t \le 0$.  Interestingly, the first moments corresponding to the EFB (solid yellow line) relax faster than those for the NEFB (dot-dashed blue line) and much faster than those for the EQ process (broken green line). This is clearly a nonlinear effect because, for $U = 0$, the EFB and NEFB share the relaxation time~\eqref{eq:relax_time0}. 
Especially for the EFB the relaxation is considerably faster for the first parameter set [panels (a)-(d)] than for the second choice [panels (e)-(h)].
This suggests that at least some intuition gained from the linear regime $U = 0$ also applies to the nonlinear dynamics. 
In accord with Figs.~\ref{fig:PDx} and \ref{fig:PD}, the position and velocity fluctuations are always smallest for the NEFB and largest for the EFB. Smaller velocity but also position variance for the EFB is obtained for the second parameter set. 

For the NEFB we were not able to analytically predict both the time evolution of the depicted variables and their asymptotic values.
To solve the full transient dynamics for the EQ and EFB is also a difficult problem. 
However, Fig.~\ref{fig:NonlinearCase} shows that the moments in question converge to the values calculated from the corresponding Boltzmann distributions \eqref{eq:px} and \eqref{eq:pv} with temperatures ${T}_0$ (EQ) and ${T}>{T}_0$ (NEQ), which provides further numerical evidence for our claims in Sec.~\ref{sec:results}.

\subsection{Heat flux}

\begin{figure}[t!]	
\centering
\includegraphics[width=0.6\columnwidth]{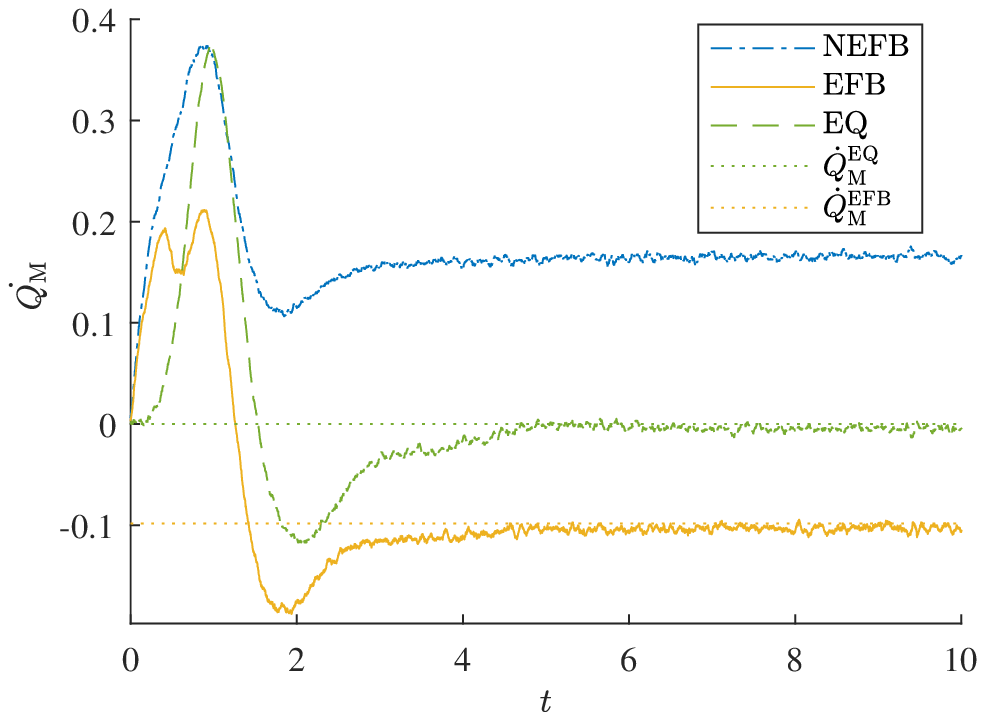}
	\caption{Heat fluxes $\dot{Q}_{\rm M}$ from the (proper) Markovian bath into the system for NEFB, EFB, and EQ processes discussed in Fig.~\ref{fig:PdfTVar} (e) -- (h). \textcolor{black}{The horizontal lines depict the stationary values of  the corresponding heat fluxes. For the EQ process, the stationary value is 0. For the EFB process, it is given by Eq.~\eqref{eq:dotQr}. For the NEFB, we have no universally valid prediction for the stationary value of $\dot{Q}_{\rm M}$. However, the small-$\tau$ expansion~\eqref{eq:smallTauv} leads to $T_{\rm eff} < T_0$ and thus it suggests a positive value of $\dot{Q}_{\rm M}$.}}
	\label{fig:dq}	
\end{figure}

Let us now investigate the heat flux $\dot{Q}_{\rm M} = \left< (- \gamma_0 v + \sqrt{2T_0\gamma_0}\xi) \dot{x} \right>$ from the ``proper'' Markovian bath into the system due to the feedback, see Tab.~\ref{tab:interpretations}C and Sec.~\ref{sec:EFB}.
In~\ref{appx:feedback_cooling}, we show that for a general EFB with a positive delay time $\tau$ the heat flux always reads
 \begin{equation}
 \dot{Q}_{\rm M}^{\rm EFB} =
 \gamma_0 \left(  
 (\sigma_v^{\rm EQ})^2 - (\sigma_v^{\rm EFB})^2\right)
 = \frac{2\gamma_0}{m }\left({T}_0-{T}\right) < 0.
 \label{eq:dotQr}
 \end{equation}
 The EFB thus always performs work on the system, which is eventually dissipated in the bath. Figure~\ref{fig:dq} displays how the heat flux in the system evolves during the relaxation processes for EQ, EFB and NEFB discussed in Fig.~\ref{fig:NonlinearCase} (e)--(h). After the initial transient period, the heat flux for EFB converges to the negative value given by Eq.~\eqref{eq:dotQr} and thus it heats both the system, as the corresponding stationary variances are larger than for the EQ process, and the proper bath. For EQ, the stationary heat flux is zero as imposed by the second law. For NEFB, the heat flux converges to a positive value. Thus the NEFB cools the system while absorbing heat from the proper bath.
 
The result~\eqref{eq:dotQr} applies for arbitrarily small positive delay $\tau$. The specific form of the feedback force~\eqref{eq:GF} allows us to also inspect what happens for vanishing delay. Then the system is still in the Boltzmann equilibrium state~\eqref{eq:px}--\eqref{eq:pv} with temperature ${T}$. However, the corresponding total noise $\eta(t) = (\sqrt{2\gamma_0{T}_0} + \sqrt{\alpha_\tau}) \xi(t)$ and friction $F_{\rm F} = -(\gamma_0 +\gamma_\tau)v(t)$ can now be interpreted as a joint influence of the standard heat bath at temperature $T_0$ and an additional `feedback heat bath' at temperature 
$T_{\rm F} = \alpha_\tau/2\gamma_{\tau} = \sqrt{{T}/{T}_0-1}\,{T}/2$. 
The laws of thermodynamics imply that heat flows from hot to cold and thus $\dot{Q}_{\rm M}$ is positive for 
$T_{\rm F}/T_0 > 1$ which occurs for $T > 2 T_0$. Further, the heat flow is zero for $T = 2 T_0$, where $T_{\rm F} = T_0$ and thus there is one global temperature only, and negative otherwise. \textcolor{black}{Evaluating the heat flux $\dot{Q}_{\rm M}^{\rm EFB}$ from Eqs.~\eqref{eq:tde2vF} and \eqref{eq:GF} with $\tau = 0$ using the approach of \ref{appx:feedback_cooling}, we find the expression
\begin{equation}
\dot{Q}_{\rm M}^{\rm EFB} = \frac{\gamma_0 T_0}{m}\sqrt{\frac{T}{T_0}-1} \left(1- \sqrt{\frac{T}{T_0}-1} \right)
\label{eq:QrEFB0}
\end{equation}
which indeed obeys the described properties.}

Since $\dot{Q}_{\rm M}^{\rm EFB}$ is strictly negative for $\tau > 0$ and can be both positive and negative for $\tau = 0$, it exhibits a discontinuity at vanishing delay, in accord with the results described in Ref.~\cite{Loos2019}. Note that the presented situation with $\tau = 0$ is physically weird since it seems impossible to record the noise and feed it back into the system without any delay. \textcolor{black}{It also yields a strange behavior as the heat flux vanishes at the point where temperatures $T_{\rm F}$ and $T_0$ are equal but $T = 2 T_0$. This means that we constructed a bath at temperature $T > T_0$ by using two strictly identical reservoirs at same the temperature $T_0$ to which the system couples via different friction coefficients. The two baths provide the same realizations of the white noise $\xi(t)$, and thus the total noise intensity is given by the sum $\sqrt{2\gamma_0T_0}+\sqrt{{\alpha_\tau}}$ of the intensities of the two noises. In contrast, connecting a system to two standard heat reservoirs always leads to equilibrium (vanishing heat flux) when the temperatures of the two baths are equal. The mathematical reason is that different reservoirs necessarily correspond to different noise realizations, regardless of their temperatures. As an example, consider heat baths A and B with friction and noise forces given by $-\gamma_{\rm A,B} v$, $\sqrt{2 \gamma_{\rm A,B} T_0} \xi_{\rm A,B}(t)$, with independent Gaussian white noises $\xi_{\rm A,B}(t)$. Then the joint action of these baths is described by the total friction $-(\gamma_{\rm A} + \gamma_{\rm B}) v$ and noise $\sqrt{2 (\gamma_{\rm A} + \gamma_{\rm B}) T_0} \xi(t)$, where $\xi(t)$ is a unit variance Gaussian white noise (correlated with $\xi_{\rm A,B}(t)$). To sum up, the formal identification of the feedback force $F_{\rm FB}$ and noise $\eta_{\rm FB}$ for $\tau=0$ as effects of a standard heat bath correctly determines the sign of the heat flux $\dot{Q}_{\rm M}$ in Eq.~\eqref{eq:QrEFB0}, but it is physically problematic.}


\subsection{Fluctuation theorems}
\label{sec:fluctuations}

\begin{figure}[t!]	
\centering
\begin{tikzpicture}
	\node (img1)  {\includegraphics[width=1.0\columnwidth]{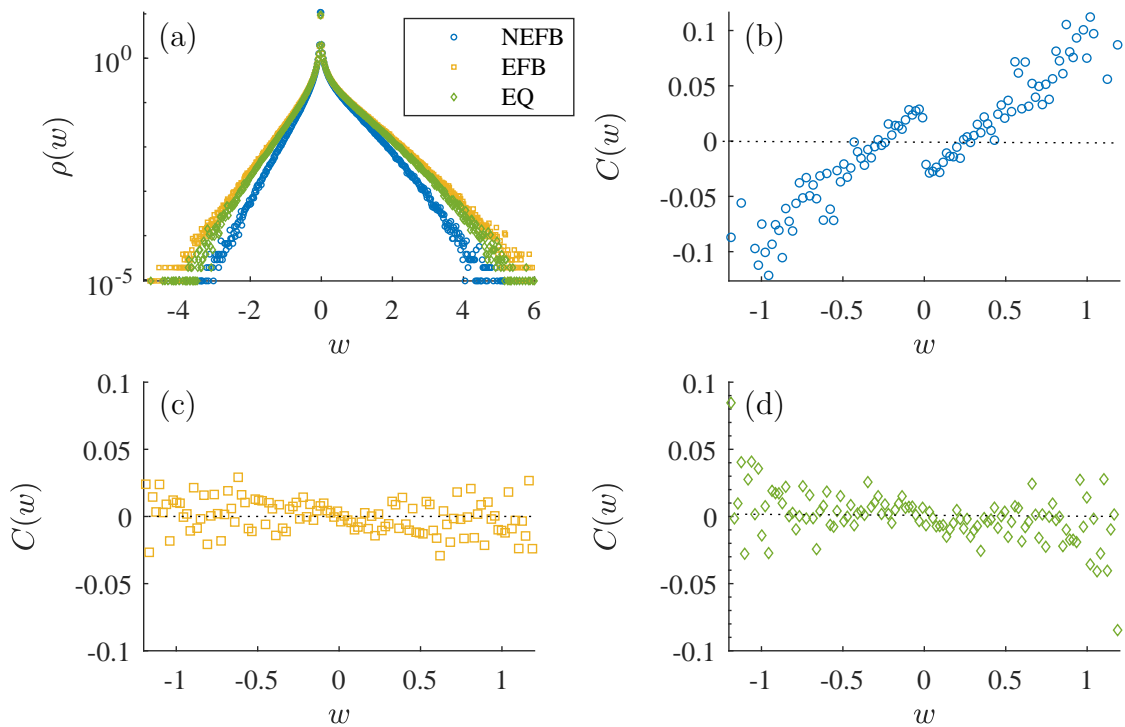}};
	\node[above=of img1, node distance=0.0cm, yshift=-2.00cm,xshift=-5.3cm] {(a)};
	\node[above=of img1, node distance=0.0cm, yshift=-2.00cm,xshift=2.5cm] {(b)};	
		\node[above=of img1, node distance=0.0cm, yshift=-6.9cm,xshift=-5.3cm] {(c)};
	\node[above=of img1, node distance=0.0cm, yshift=-6.9cm,xshift=2.5cm] {(d)};	
\end{tikzpicture}
	\caption{Test of the Crooks fluctuation theorem. (a) Probability densities for work for NEFB, EFB and EQ processes departing from the stationary state of Fig.~\ref{fig:PdfTVar} and driven with a time symmetric protocol~\eqref{eq:k6prot} for the potential~\eqref{eq:Unonl}. The remaining panels show the function~\eqref{eq:CrooksC} for NEFB (b), EFB (c), and EQ (d). The data suggest that the theorem holds for EFB and EQ, where $C(w) \approx 0$. The shown results were obtained from $5\times 10^6$ runs of BD simulation with time-step $dt = 10^{-3}$.
 Except for the time-dependent driving, all parameters are the same as in Fig.~\ref{fig:PdfTVar}.}
	\label{fig:rhow}	
\end{figure}

We conclude the numerical part of the paper by testing the work fluctuation theorems~\eqref{eq:Jarzynski} and \eqref{eq:Crooks}. To this end, we let the system relax into the steady state corresponding to the parameter regime of Fig.~\ref{fig:PdfTVar} (e) -- (h) and then we switch on the time-symmetric protocol
\begin{equation}
k_6 = 1 + 0.9\sin(\pi t)   
\label{eq:k6prot}
\end{equation}
with $t \in (0,1)$ for the potential~\eqref{eq:Unonl}. During the time-dependent driving, we measure the stochastic work
\begin{equation}
    w = \int_0^1 dt\, \frac{\partial{U}[x(t),t]}{\partial t} = \int_0^1 dt\, \dot{k}_6 x^6(t)/6
\end{equation}
and sample its PDF $\rho(w)$. Due to the symmetry of the protocol, the time-reversed process (R) and the forward process (F) in the fluctuation theorems~\eqref{eq:Jarzynski} and \eqref{eq:Crooks} coincide and the free energy difference $\Delta F$ vanishes. Validity of the Crooks fluctuation theorem~\eqref{eq:Crooks} for the acquired PDFs thus implies that 
\begin{equation}
  C(w) \equiv  \log \left[ \frac{\rho(w)}{\rho(-w)} \exp(w/{T}) \right] = 0.
  \label{eq:CrooksC}
\end{equation}
In Fig.~\ref{fig:rhow}~(a) we show the resulting PDFs for work obtained for the NEFB, EFB, and EQ processes. The panels (b) -- (d) then show that from the three processes only the EFB (c) and EQ (d) yield $C(w) = 0$ and thus fulfill the Crooks fluctuation theorem~\eqref{eq:CrooksC}.

\begin{figure}[t!]	
\centering
\includegraphics[width=0.6\columnwidth]{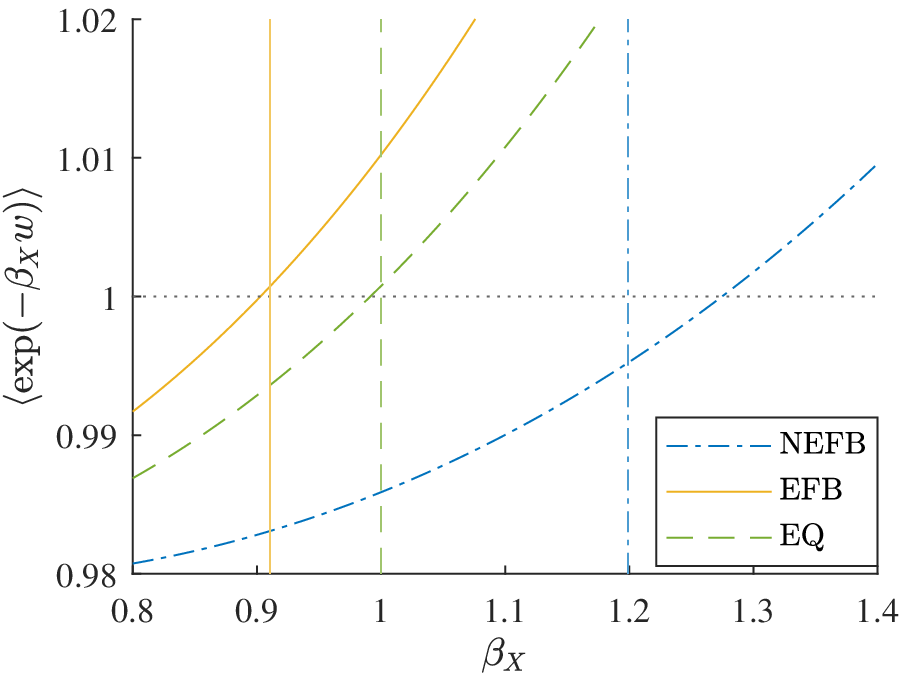}
	\caption{Test of the Jarzynski equality for work PDFs from Fig.~\ref{fig:rhow}. The vertical lines for the individual processes correspond to temperatures evaluated from the stationary variance of the velocity as $1/\beta_X = m \left<v^2\right>$ assuming equipartition theorem. For the EFB and EQ this temperature equals to the temperature measured in any other way (e.g., from position or velocity PDF). The NEFB in general induces a non-equilibrium steady state and thus corresponding temperatures measured in different ways are in general different~\cite{Cugliandolo2011,Holubec2020}.}
	\label{fig:Jarzynski}	
\end{figure}

The validity of the Jarzynski equality is tested in Fig.~\ref{fig:Jarzynski}, where we show values of averages $\left< \exp(- \beta_{\rm X} w)\right>$ over the sample PDFs for work as functions of the parameter $\beta_{\rm X}$. For EFB and EQ, we find that the average equals to one for $\beta_{\rm X} = {1/T}$ and $1/{T}_0$, respectively, proving the validity of Jarzynski equality~\eqref{eq:Jarzynski} with $\Delta F = 0$ in these cases. For the NEFB, the system starts out of equilibrium so that it is not clear which (inverse) temperature should be used in~Eq.~\eqref{eq:Jarzynski}. In the figure, we at least tested that choosing the temperature obtained form the variance of the velocity, $1/{\beta}_{\rm X} = 2\sigma_v^2$, does not yield $\left< \exp(- \beta_{\rm X} w)\right> = 1$.

\section{Beyond equilibrium feedback}
\label{sec:response}

In this section, we discuss possible analytical extensions of the equilibrium delay processes that might help to better understand general delay processes.

\subsection{Classical linear response theory}
\label{sec:clresponse}

Any Langevin equation where the friction and noise obey the FDR~\eqref{eq:FDTGen} can be 
thought of as a result of coarse-graining the full set of Hamiltonian equations for the system of interest and the corresponding bath over the bath degrees of freedom. This means that, the system with equilibrium delay can be regarded as a standard Hamiltonian system, which implies applicability of the classical linear response theory~\cite{Kubo1966,Kubo2012,Zwanzig2001}. It states that  the time-evolution of the mean value $\left< A(t) \right>$ induced by perturbations of the equilibrium system with Hamiltonian $H= U + mv^2/2$ in the form $H + \varepsilon f(t) B$ starting at time $t=0$ reads~\cite{Zwanzig2001}
\begin{equation}
\left<A(t) \right>_1 = \left<A(t) \right>_0 + \frac{\varepsilon}{T} \int_0^t ds\, f(s) \left<A(t-s) \dot{B}(0)\right>_0.
\label{eq:lin_resp}
\end{equation}
One assumes that the averages in the perturbed system can be expanded as $\left<\dots\right> = \left<\dots\right>_0 + \varepsilon \left<\dots\right>_1 + \dots$, where the subscript 0 denotes average taken over the unperturbed Boltzmann PDF corresponding to Hamiltonian $H$~\eqref{eq:px} and \eqref{eq:pv}, the subscript 1 denotes averages taken over the exact PDF up to the order $\varepsilon$, and so on.

\begin{figure}[t!]	
\begin{tikzpicture}
	\node (img1) {\includegraphics[trim=0 20 0 0, width=1.0\columnwidth]{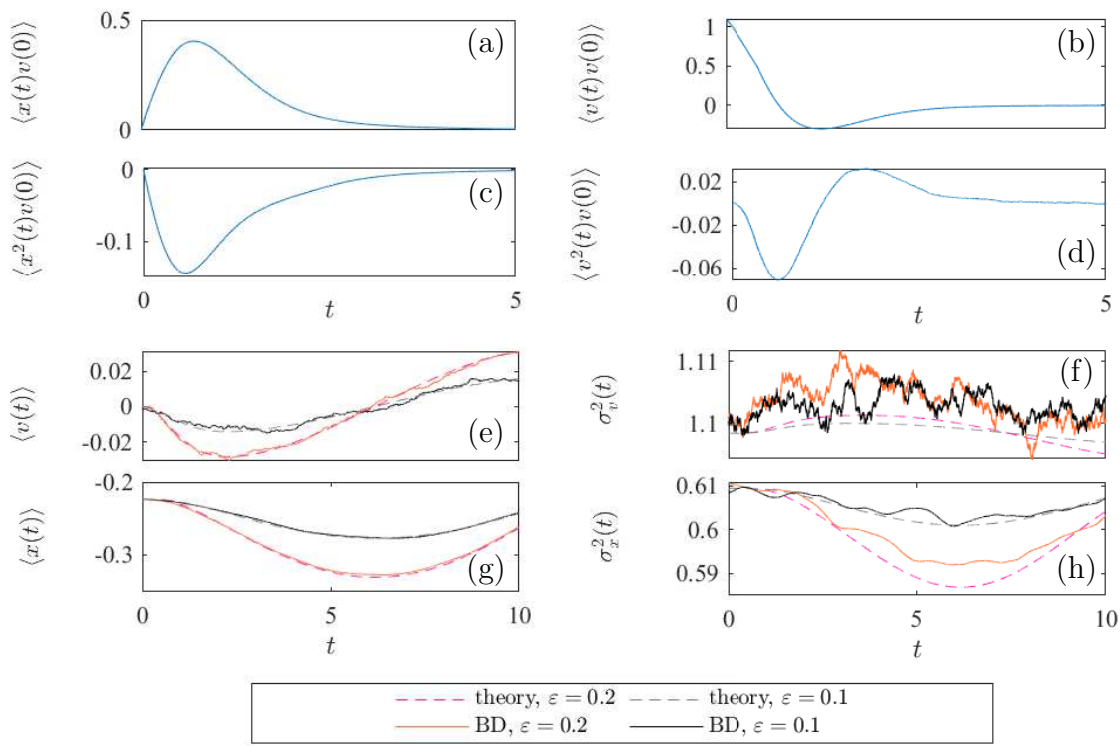}};
	\node[above=of img1, node distance=0.0cm, yshift=-2.1cm,xshift=-1.15cm] {(a)};
	\node[above=of img1, node distance=0.0cm, yshift=-2.1cm,xshift=6.7cm] {(b)};	
		\node[above=of img1, node distance=0.0cm, yshift=-4.1cm,xshift=-1.15cm] {(c)};
	\node[above=of img1, node distance=0.0cm, yshift=-4.9cm,xshift=6.7cm] {(d)};	
	\node[above=of img1, node distance=0.0cm, yshift=-7.3cm,xshift=-1.15cm] {(e)};
	\node[above=of img1, node distance=0.0cm, yshift=-6.48cm,xshift=6.7cm] {(f)};
	\node[above=of img1, node distance=0.0cm, yshift=-9.1cm,xshift=-1.15cm] {(g)};
	\node[above=of img1, node distance=0.0cm, yshift=-9.1cm,xshift=6.7cm] {(h)};
\end{tikzpicture}
	\caption{Linear response theory: (a)-(d) response functions from BD simulations used to evaluate the time evolution of the first two central moments of velocity and position (e)-(h) in the equilibrium system of Fig.~\ref{fig:PdfTVar} perturbed by the time-dependent force~\eqref{eq:forceLR}. As expected, agreement between the approximate theory (dashed lines) and simulation (noisy solid lines) is better for smaller $\epsilon = 0.1$ (black and gray lines) than for the larger one $\epsilon = 0.2$ (pink and orange lines).}
	\label{fig:LRT}	
\end{figure}

We test the linear response theory using the specific equilibrium velocity feedback system discussed in Sec.~\ref{sec:discussionA}. We perturb the Hamiltonian by the term $\varepsilon f(t) x$. This term corresponds to a homogeneous time-dependent force $- \varepsilon f(t)$ and thus the dynamical equation for velocity reads
\begin{equation}
m\dot{v}(t) =  -\frac{\partial{U}}{\partial{x}}  + F_{\rm FB} - \gamma_0 v(t) + \sqrt{2T_0\gamma_0}\xi(t)
 -\varepsilon f(t).
\end{equation}
The potential $U$ is given by Eq.~\eqref{eq:Unonl} and the feedback force by Eq.~\eqref{eq:GF}. We consider the parameter regime of Fig.~\ref{fig:PdfTVar} and the specific perturbation
\begin{eqnarray}
\varepsilon f(t)= \varepsilon \sin(\pi t/10).
\label{eq:forceLR}
\end{eqnarray}
In Figs.~\ref{fig:LRT} (a)-(d) we show the time correlation functions $\left<A(t)\dot{x}(0)\right>_0 = \left<A(t)v(0)\right>_0$, $A = x, v, x^2, v^2$ obtained using BD simulations of this system with $\varepsilon = 0$. The first two central moments of velocity and position obtained using Eq.~\eqref{eq:lin_resp} with force~\eqref{eq:forceLR} via these time correlation functions are depicted in Figs.~\ref{fig:LRT} (e)-(h) together with the corresponding quantities obtained from BD simulation of the perturbed system. The figures show good agreement between simulations and linear response theory, which improves for smaller $\varepsilon$.

The validity of the linear response theory can be rationalized as follows. Even though we do not have an exact dynamical equation for the PDF for the equilibrium delay system, we know that the PDF for system and bath obeys a Liouville equation. The corresponding Liouville operator is composed of the system Hamiltonian $H$, the bath Hamiltonian, and the system-bath interaction energy. Even though it is hard to identify the latter two, one can rely on this Liouville equation as a starting point for perturbation theories around the parameter regime of the equilibrium delay. \textcolor{black}{For example, one can derive Eq.~\eqref{eq:lin_resp} using the textbook approach of Ref.~\cite{Zwanzig2001}}.

\subsection{Langevin equation}
\label{sec:AppFromLangevin}

The \textcolor{black}{classical linear response theory~\eqref{eq:lin_resp} applies} only for perturbations that can be subsumed into the Hamiltonian \textcolor{black}{of the system}. Other perturbations  \textcolor{black}{can be treated, e.g., on the footing of linear irreversible thermodynamics~\cite{callen1985} or directly on the level of the Langevin equations~\eqref{eq:tde2x} and \eqref{eq:tde2v}. In order to present two simple examples of the latter type}, we write these equations in the form of the equilibrium interpretation of Tab.~\ref{tab:interpretations}B
\begin{equation}
    \dot{x} = v, \quad m\dot{v} = \left(-\frac{\partial U}{\partial x} + F_{\rm F} + \eta \right) + \varepsilon g,
    \label{eq:vexp}
\end{equation}
where the term proportional to $\varepsilon$ is a perturbation. Perturbations dependent on time and/or time-delayed variables again require evaluation of time-correlation functions, which can rarely be obtained analytically. Perturbations that depend only on position can be absorbed into the potential, and the stationary PDF can be evaluated exactly. To obtain non-trivial analytical results, we will investigate two properties of steady states induced by perturbations of the form $g = g[v(t)]$, i.e. which depend solely on velocity. \textcolor{black}{However, the obtained general restrictions \eqref{eq:VirialE1} and \eqref{eq:48} on the system dynamics apply for an arbitrary function $g$. In particular, $g$ can, for these expressions, be a nonlinear function of position and velocity in the past. In such a case, the resulting process~\eqref{eq:vexp} is a truly nonlinear delay differential equation.}

\emph{Virial theorem:} The viral theorem states that twice the average kinetic energy of a system equals to the virial $-\left<F x\right>$~\cite{goldstein2002}. For the system at hand, the total force $F$ is given by the right-hand side (R.H.S.) of Eq.~\eqref{eq:vexp}. For the unperturbed system, this implies that
\begin{equation}
m \left<v^2\right> = -\left< \left(-\frac{\partial U}{\partial x} + F_{\rm F} + \eta \right) x\right>
\label{eq:VirialE0}
\end{equation}
as can be checked directly from the Langevin equations after multiplying Eq.~\eqref{eq:tde2x} by $x$, Eq.~\eqref{eq:vexp} with $\varepsilon = 0$ by $v$, summing the result, and taking the stationary ensemble average so that the time derivative of the cross-correlation $\left<xv\right>$ vanishes. Repeating this procedure for nonzero perturbation in Eq.~\eqref{eq:vexp}, we find that if $\left<g(v) \right>_0 = 0$ the unperturbed virtial theorem~\eqref{eq:VirialE0} remains valid up to first order in $\varepsilon$ because $\left<x g(v) \right>_0 = \left<x\right>_0 \left<g(v) \right>_0$. More generally, we find that
\begin{equation}
 m \left<v^2\right>_n + \left< \left(-\frac{\partial U}{\partial x} + F_{\rm F} + \eta \right) x\right>_n = - \varepsilon \left<x g(v) \right>_{n-1}
\label{eq:VirialE1}
\end{equation}
holds for all corrections of order $n\ge 1$. Even though this result cannot be evaluated explicitly for $n>1$ since $ \varepsilon^{n}  \left<x g(v) \right>_1$ is unknown, it can provide a stringent consistency check for simulations.

\begin{figure}[t!]	
\centering
\begin{tikzpicture}
	\node (img1) {\includegraphics[trim=0 20 0 0, width=1.0\columnwidth]{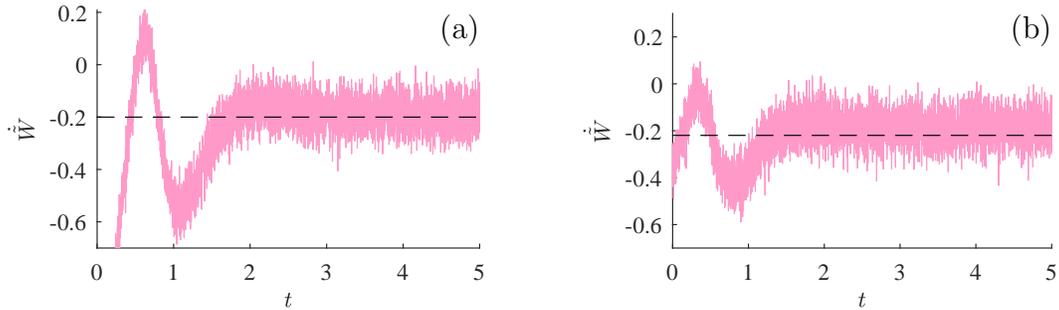}};
	\node[above=of img1, node distance=0.0cm, yshift=-2.30cm,xshift=-0.6cm] {(a)};
	\node[above=of img1, node distance=0.0cm, yshift=-2.30cm,xshift=7.0cm] {(b)};	
\end{tikzpicture}
	\caption{Test of the explicit result~\eqref{eq:Papprox} (dashed lines) against the simulation (solid line). (a) $\varepsilon = 0.2$ and $n=1$. (b) $\varepsilon = 0.01$ and $n=3$. Other parameters are the same as in Fig.~\ref{fig:PdfTVar}.}
	\label{fig:PR}	
\end{figure}

\emph{Power:} The power $\left<F v\right>$ exerted by the total force $F$ on the R.H.S. of Eq.~\eqref{eq:vexp} vanishes in the steady state of the unperturbed system since the time-derivative of $\left<v^2\right>$ vanishes. With the perturbation switched on, we find
\begin{equation}
  \left< \left(-\frac{\partial U}{\partial x} + F_{\rm F} + \eta \right) v\right>_n =
  - \varepsilon \left<v g(v)\right>_{n-1}.
  \label{eq:48}
\end{equation} 
Besides providing another set of expressions useful as consistency checks in simulations, this equation provides an explicit non-trivial result for $n=1$. Then the R.H.S. $- \varepsilon \left<v g(v)\right>_{0}$ is in general non-zero and can be evaluated as average over the PDF~\eqref{eq:pv}. For example, for $g(v) = v$, we find
\begin{equation}
  \dot{\tilde{W}} \equiv \left< \left(-\frac{\partial U}{\partial x} + F_{\rm F} + \eta \right) v\right>_n =
  - \varepsilon T.
  \label{eq:Papprox}
\end{equation}
Figure~\ref{fig:PR} verifies this approximate result by BD simulations for $n=1$ and $n=3$ and the same unperturbed dynamics as in Sec.~\ref{sec:clresponse} above.

\section{Conclusion}
\label{sec:conclusion}

Using the second FDR, we have identified situations where one can interpret time-delayed feedback forces proportional to velocity or position in Langevin equations as friction forces imposed by an equilibrium bath. Our analysis reveals a previously unnoticed class of nonlinear stochastic delay differential equations, which can be treated analytically. They describe processes that obey standard thermodynamic constraints. In particular, their long-time distributions are of Gibbs canonical form and they obey standard fluctuation theorems. From the point of view of control theory, especially passivity-based control, the corresponding equilibrium feedback processes are automatically stable and passive. However, their dynamics retains the full complexity of generic delay processes. One disadvantage is that the equilibrium feedback always heats up the system and thus, unlike generic feedback protocols, cannot be used for standard feedback cooling.

As a practical demonstration, we have realized the equilibrium velocity feedback using Brownian dynamics simulations and shown that it exhibits all the formally derived properties. For so-called equilibrium velocity feedback, not only the velocity at an earlier time but also the noise at that time are used to drive the system at present. \textcolor{black}{Such feedback can nowadays be realized in practice in feedback experiments~\cite{Gernert2016} with Brownian particles \cite{Baraban2013,Qian2013,Bregulla2014,Khadka2018,Lavergne2019,Bauerle2020,Penny2021} or robots~\cite{Mijalkov2016,Leyman2018,Piwowarczyk2019}. In particular, the so-called velocity damping protocols for feedback cooling of trapped microscopic particles record the particle velocity and later, after an experimental latency, apply to the particle a force proportional to that velocity~\cite{Ferialdi2019,Penny2021}. Assuming that the position and velocity dependence of the systematic force in the dynamical equation for particle motion is known, measuring both its velocity and position at time $t$ allows to determine the thermal noise, which can then be applied to the particle in the future, similarly as the velocity-dependent force. The potential drawback of this approach is that the measured noise will be affected by measurement uncertainties and finite measurement time resolution. The latter means that the obtained noise will effectively be integrated (low-pass filtered) over one measurement frame, which is however also the case in our BD simulations. Other promising setups, where the equilibrium feedback might be realised are state-of-the-art bath engineering experiments~\cite{Murch2012}. Finally, equilibrium feedback of the same type as in our computer simulations can be realised in experimental setups where artificial noise completely overshadows thermal noise. An example is shaken granular matter~\cite{DAnna2003}, where the noise is realised by shaking the granular system.}

From a theoretical perspective, we believe that our results will shed further light on investigations of the dynamics and thermodynamics of nonlinear stochastic delay differential equations, which are known to be immensely resistant to analytical treatments. For example, the known stationary distributions for the equilibrium delay processes can serve as starting points of new perturbation theories valid for arbitrarily large delays. And, the known thermodynamics of the equilibrium delay processes can help to better understand the individual contributions to the total entropy production derived for nonlinear stochastic delay differential equations as studied in Refs.~\cite{Munakata2014,Rosinberg2015,Rosinberg2017}. 

\section*{Acknowledgments}
We acknowledge funding through a DFG-GACR cooperation by the Deutsche Forschungsgemeinschaft (DFG Project No 432421051) and by the Czech Science Foundation (GACR Project No 20-02955J). VH gratefully acknowledges support by Humboldt foundation.

\section*{References}

\bibliographystyle{iopart-num}
\bibliography{References}

\appendix


\section{No feedback cooling with EQ feedback}
\label{appx:feedback_cooling}

Consider the heat flux $\dot{Q}_{\rm M}^{\rm EFB}$ from the proper bath into the system in the steady state created by an equilibrium feedback. The force applied by the proper bath on the system is $-\gamma_0 v(t) + \sqrt{2 T_0 \gamma_0}\xi(t)$ and thus
\begin{align}
    \dot{Q}_{\rm M}^{\rm EFB}  = & - \gamma_0\langle  v^2 \rangle
+ \sqrt{2T_0\gamma_0}\,
\langle \xi(t) v(t) \rangle
\end{align}
with $ \langle v^2\rangle = \lim_{t\to\infty} \langle v(t)^2\rangle =  \sigma^{\rm EFB}_v = {T}/ m$.
To calculate $\langle \xi(t) v(t) \rangle$, we use the formal solution
\begin{equation}
    v(t) = \frac{1}{m} \int_0^t dt' \bigg[F[x(t'),v(t'),t'] + F_{\rm D}[x(t'-\tau),v(t'-\tau)] + \sqrt{2 T_0 \gamma_0}\xi(t') + \eta_{\rm FB}(t')\bigg].
\end{equation}
For a positive delay, $\tau > 0$, the causality implies that all terms on the R.H.S. except for $\xi(t')$ are independent of the white noise $\xi(t)$ at time $t$ (values of velocity and position at time $t' \le t$ were not affected by the white noise yet, and it is also reasonable to assume that the feedback noise $\eta_{\rm FB}(t')$ can not be constructed in such a way that it would depend on $\xi(t)$). In symbols we obtain
\begin{equation}
\langle \xi(t) v(t) \rangle = \frac{1}{m} \int_0^t   dt' 
\sqrt{2 T_0 \gamma_0} \delta(t-t')
 =  \frac{1}{m} \sqrt{\frac{ T_0 \gamma_0}{2}}.
\label{eq:xiv}
\end{equation}
Using $\sigma^{\rm EQ}_v = {T}_0/m$ we get the stationary flux $\dot{Q}_{\rm M}$ induced by the EQ feedback in the form
 \begin{equation}
 \dot{Q}_{\rm M}^{\rm EFB} =
 \gamma_0 \left(  
 (\sigma_v^{\rm EQ})^2 - (\sigma_v^{\rm EFB})^2\right)
 = \frac{2\gamma_0}{m}\left(T_0-T\right).
 \end{equation}
As might have been anticipated from the beginning, the EQ feedback can cool the proper bath ($\dot{Q}_{\rm M}^{\rm EFB}>0$) only if it leads to a smaller velocity variance \textcolor{black}{(effective temperature)} than the proper bath. Let us now show that this can happen only if the feedback force contains also a time-local term in velocity.

Inserting the total noise $\eta(t) = \sqrt{2T_0\gamma_0}\xi(t) + \eta_{\rm FB}(t)$ (see Tab.~\ref{tab:interpretations}) into the FDR~\eqref{eq:FDTGen}, we find
\begin{equation}
\left<\eta(t)\eta(t')\right>/{T}
= 2(\gamma_0 +\delta) \delta(t-t') + \dots = (2T_0\gamma_0 + \epsilon)/{T}\, \delta(t-t') + \dots.
\label{eq:TCFApp1}
\end{equation}
The first line corresponds to the time-local component of the friction kernel. Specifically, the term proportional to $\gamma_0$ corresponds to the background friction $\gamma_0 v(t)$. And the term proportional to $\delta$ stems from the time-local component of the feedback force. The remaining terms abbreviated by $\dots$ are determined by time non-local components of the friction kernel. The second line corresponds to the actual noise correlations. The term proportional to $2T_0\gamma_0$ is obtained from the background noise and the term proportional to $\epsilon \delta(t-t')$ originates from the time-local ($t-t' = 0$) component of $\left<\eta_{\rm FB}(t)\eta_{\rm FB}(t')\right>$. The remaining terms are given by the cross correlations $\left<\xi(t)\eta_{\rm FB}(t')\right>$ and the $t-t' \neq 0$ component of $\left<\eta_{\rm FB}(t)\eta_{\rm FB}(t')\right>$. 

Demanding that prefactors in front of the $\delta$-functions in Eq.~\eqref{eq:TCFApp1} equal, we find
\begin{equation}
\frac{{T}_0}{{T}} = \frac{2(\gamma_0 + \delta) - \epsilon/T}{2\gamma_0}.
\end{equation}
Since $\epsilon$ is the variance of the feedback noise $\eta_{\rm FB}(t)$, it must be positive. The feedback can thus cool the system and the ambient bath ($T<T_0$ and  $\dot{Q}_{\rm M}^{\rm EFB} > 0$) only if it contains a strong enough time-local component of the friction $-\delta v(t)$, $\delta > \epsilon/2T$. Otherwise ${T}\geq {T}_0$ and $\dot{Q}_{\rm M}^{\rm EFB} \le 0$ and thus the cooling by equilibrium feedback is not possible. Realizing a feedback force containing a term proportional to a non-delayed velocity of the system seems technologically impossible and thus we conclude that, unlike generic feedback, the equilibrium feedback cannot be used to cool the system in practical setups~\cite{Bushev2006,Goldwater2019}.

\section{Noise generation in practice}
\label{appx:noises_in_reality}

An arbitrary Gaussian noise with a given positive power spectrum, and thus also any noise $\eta(t)$ fulfilling the conditions~\eqref{eq:FDTGen} and \eqref{eq:PowerSpectrumGen}, can be realized in practice using one of the standard procedures for generating a Gaussian process with given autocorrelation function~\cite{Shinozuka1991,Pichot2016,Graham2018}.

Concerning the position delay of Sec.~\ref{sec:PF} in the parameter regime~\eqref{eq:inequality_x}, our brute-force implementation of the spectral method using Eq.~(35) in Ref.~\cite{Shinozuka1991} does not yield satisfactory results. However, we can recommended the method of Refs.~\cite{Pichot2016,Graham2018} based on a discrete representation of the noise $\eta(t)$ by $M_t {\mathbf e}$, where ${\mathbf e}$ is a column vector of independent Gaussian white noises and the matrix $M_t$ is given by a square root of a matrix describing the autocorrelation function~\eqref{eq:noiseTCFxv}.

A noise fulfilling the conditions \eqref{eq:noiseTCFv} and \eqref{eq:inequalities_gamma} corresponding to the velocity delay of Sec.~\ref{sec:VF} can be constructed analytically by introducing the ansatz 
\begin{equation}
\eta(t) = \sqrt{\alpha_0} \xi(t) + \sqrt{\alpha_\tau} \xi(t-\tau).
\label{eq:etaDelayEq}
\end{equation}
Here, $\alpha_0 = 2{T}_0 \gamma_0 $, $\xi(t)$ is the zero-mean, unit-variance, Gaussian white noise, i.e. $\left<\xi(t)\right> = 0$, $\left<\xi(t)\xi(t')\right> = \delta(t-t')$. For $\tau > 0$, such $\eta(t)$ obeys Eq.~\eqref{eq:noiseTCFv} if $(\alpha_0 + \alpha_\tau)/{T} = 2  \gamma_0$ and $ \sqrt{\alpha_0 \alpha_\tau}/{T} = \gamma_\tau$. Solving these equations with respect to $\alpha_0$ and $\alpha_\tau$, we obtain the two roots
\begin{equation}
\alpha_0/{T}  = \gamma_0 \pm \sqrt{\gamma_0^2-\gamma_\tau^2}, \quad \alpha_\tau /{T} = \gamma_0 \mp \sqrt{\gamma_0^2-\gamma_\tau^2},
\label{eq:alpha0}
\end{equation}
which are equivalent due to the symmetry $\alpha_0 \leftrightarrow \alpha_\tau$ of the noise leading to the same dynamics of $v(t)$. In agreement with our discussion of the power spectrum~\eqref{eq:PSv}, this mapping breaks down if $\gamma_0 < |\gamma_\tau|$ when $\eta(t)$ becomes complex. 

In standard Brownian dynamics simulations, the stochastic differential equation $\dot{v}(t) = f(t,t-\tau) + \sigma \xi(t)$ is usually solved using the Euler method leading to the update rule $v(t+dt) = v(t) + f(t,t-\tau)dt + \sigma \sqrt{dt} L(t)$, where $\sqrt{dt} L(t) = \int_t^{t+dt}dt' \xi(t')$ is a zero-mean Gaussian random variable with variance $dt$. The equilibrium velocity delay process \eqref{sec:VF} with the noise \eqref{eq:etaDelayEq}
has the form $\dot{v}(t) = f(t,t-\tau) + \sqrt{\alpha_0} \xi(t) + \sqrt{\alpha_\tau} \xi(t-\tau)$. And it can be simulated using the update rule $v(t+dt) = v(t) + f(t,t-\tau)dt + \sqrt{\alpha_0 dt}L(t)  + \sqrt{\alpha_\tau dt} L(t-\tau)$. To simulate the process, one thus has to keep track of the noise sequence $L(t')$ for $t' \in [t-\tau,t]$.

\section{Parameter sets used in Figs.~\ref{fig:PdfT} and \ref{fig:PdfTVar}}
\label{appx:parameterChoice}
The two parameter sets considered in Figs.~\ref{fig:PdfT} and \ref{fig:PdfTVar} are chosen as follows. The first one $(\tilde{\gamma}_0 \tau,\tilde{\gamma}_\tau \tau) \approx (0.28,0.28)$ minimizes the ratio
\begin{equation}
\frac{t_{\rm R}}{t_{\rm R}^{\rm EQ}} = \frac{\tilde{\gamma}_0 \tau}{\Re\left[\tilde{\gamma}_0 \tau-W\left(-e^{\tilde{\gamma}_0 \tau} \tilde{\gamma}_\tau \tau\right)\right]}
\label{eq:tEqtotR}
\end{equation}
of the relaxation time $t_{\rm R}^{\rm EQ} \equiv 1/\tilde{\gamma}_0$ for the EQ process and $t_{\rm R}$~\eqref{eq:relax_time0} for the EFB and NEFB for $U = 0$. 
The second one $(\tilde{\gamma}_0 \tau,\tilde{\gamma}_\tau \tau) \approx (0.42,0.24)$ minimizes the measure 
\begin{equation}
\frac{t_{\rm R}}{t_{\rm R}^{\rm EQ}} \left(\frac{\sigma_v^{\rm EFB}}{\sigma^{\rm NEFB}_v}\right)^2
\label{eq:tEqtotRvar}
\end{equation}
of the trade-off
between the relaxation time ratio~\eqref{eq:tEqtotR} and the ratio of stationary velocity variance $(\sigma_v)^2 = \left< v^2\right>$ for EFB and NEFB. For NEFB we take the variance for $U=0$, when it can be calculated analytically as
\begin{equation}
\left(\sigma_v^{\rm NEFB}\right)^2 \equiv \sigma_v^2 = \frac{ T_0 }{m}\frac{\tilde{\gamma}_0[\Omega+\tilde{\gamma}_\tau \sinh(\Omega\tau)]}{\Omega\left[\tilde{\gamma}_0 + \tilde{\gamma}_\tau \cosh(\Omega\tau)\right]},
\label{eq:variancefeedtext}
\end{equation}
where $\Omega = \sqrt{\tilde{\gamma}_0^2 - \tilde{\gamma}_\tau^2}$. For EFB, the variance $\left(\sigma_v^{\rm EFB}\right)^2 = {T}/m$ follows for any $U$ from equipartition. 

In addition, we also tested the parameter set $(\tilde{\gamma}_0 \tau,\tilde{\gamma}_\tau \tau) \approx (0.34,0.034)$ which minimizes the ratio $\sigma_v^{\rm EFB}/\sigma^{\rm NEFB}_v$ but due to the small magnitude of the feedback force the numerical results for NEFB, EFB, and EQ are hardly distinguishable and we decided not to show them. 

\section{Derivation of Eq.~\eqref{eq:variancefeedtext}}
\label{appx:variance}
Consider the simple velocity process
\begin{equation}
\dot{v}(t) =  - \tilde{\gamma}_0 v(t) - \tilde{\gamma}_\tau v(t-\tau) + \sqrt{\alpha} \xi(t).
\label{eq:tdeApp}
\end{equation} 
In the steady state, $\left<v\right> = 0$ and the variance $\sigma^2_v = \left<v^2\right>$ can be evaluated as follows.

The general solution to Eq.~\eqref{eq:tdeApp} reads~\cite{Geiss2019}
\begin{equation}
v(t) = \lambda(t) v_0 - \tilde{\gamma}_{\tau}\int_{-\tau}^0 dt'\, \lambda(t-t'-\tau)  v(t') + \sqrt{\alpha} \int_0^t dt' \, \lambda(t-t') \xi(t'),
\label{eq:gen_sol_phi}
\end{equation}
where $v_0 = v(0)$. The Green's function $\lambda(t)$ solves Eq.~\eqref{eq:tdeApp} with vanishing noise term ($\alpha =0$), i.e.
\begin{equation}
\dot{\lambda}(t) = - \tilde{\gamma}_0 \lambda(t) - \tilde{\gamma}_{\tau} \lambda(t-\tau),
\label{eq:lambdaTEQ}
\end{equation}
and the initial condition $\lambda(t<0) = 0$ and $\lambda(0) = 1$.
The most straightforward way for finding $\lambda(t)$ is to employ a Laplace transformation in time. The result is~\cite{McKetterick2014}
\begin{equation}
\lambda(t) = \sum_{l=0}^{\infty}
\frac{(- \tilde{\gamma}_{\tau})^l}{l!}\left(t - l\tau\right)^l {\rm e}^{- \tilde{\gamma}_0 \left(t - l\tau \right)} \Theta\left(t - l\tau\right).
\end{equation}
In the stable regime, $t_{\rm R} > 0$ (cf Eq.~\eqref{eq:relax_time0}), where the Green's function eventually decays to zero, $\lim_{t\to\infty}\lambda(t) = 0$, the general solution~\eqref{eq:gen_sol_phi} to Eq.~\eqref{eq:tdeApp} can be used for calculation of the time-correlation function, $C(t) = \lim_{t_0\to \infty}\left<v(t_0+t)v(t_0)\right>$. The stationary variance $\sigma_v^2 = C(0)$, we are actually interested in, comes as a by-product of this calculation. 

We find that for $t>0$ (see also Ref.~\cite{Geiss2019})
\begin{equation}
C(t) = \alpha \lim_{t_0\to \infty}
\int_0^{t_0} dt'\, \lambda(t+t_0-t')\lambda(t_0-t').
\label{eq:Ckt_easy}
\end{equation}
This expression can already be used for plotting the time correlation function, however, it is not very suitable for inferring its properties. Following the approach in Refs.~\cite{Frank2003,Geiss2019}, we take the time derivative of Eq.~\eqref{eq:Ckt_easy} and use Eq.~\eqref{eq:lambdaTEQ} for the Green's function to obtain the dynamical equation
\begin{equation}
\dot{C}(t) = - \tilde{\gamma}_{\tau} C(t-\tau) - \tilde{\gamma}_0 C(t)
\label{eq:dCkt}
\end{equation}
valid for $t>0$ due to the nonanalyticity of $\lambda(t)$ at $t=0$. The solution to this equation is given by Eq.~\eqref{eq:gen_sol_phi}:
\begin{equation}
C(t) = \lambda(t) C_0 - \tilde{\gamma}_{\tau}\int_{-\tau}^0 dt'\, \lambda(t-t'-\tau)  C(t')
\label{eq:Cktformal}
\end{equation}
and thus the decay time of the time-correlation function is given by the decay time~\eqref{eq:relax_time0} of the Green's function $\lambda(t)$. To evaluate the above expression, we need to find the stationary variance $\alpha_0 \equiv C(0)$ and the delayed initial condition $C(t)$ for $t \in (-\tau,0)$. This can be done as follows. Employing the symmetry $C(t) = C(-t)$ of the stationary time-correlation function, we rewrite Eq.~\eqref{eq:dCkt} as
\begin{equation}
\dot{C}(t) = - \tilde{\gamma}_{\tau} C(\tau-t) - \tilde{\gamma}_0 C(t).
\label{eq:dCkt2}
\end{equation}
For $t \in (0,\tau)$, we can differentiate this equation once again. The result is
\begin{equation}
\ddot{C}(t) = \Omega^2 C(t),
\label{eq:ddCkt}
\end{equation}
where we used Eq.~\eqref{eq:dCkt} and defined the (possibly imaginary) frequency $\Omega =  \sqrt{\tilde{\gamma}_0^2 - \tilde{\gamma}_\tau^2}$. From Eq.~\eqref{eq:ddCkt}, we find that for $t\in [-\tau,\tau]$
\begin{equation}
C(t) = C_0 \cosh\left(\Omega t\right) +
\dot{C}_0 \Omega^{-1} \sinh\left(\Omega |t|\right).
\label{eq:Cktini}
\end{equation}
Here, $\dot{C}_0 = \lim_{t\to 0+} \dot{C}(t)$ denotes the time-derivative of the time-correlation function, that is discontinuous at $t=0$~\cite{Frank2003}, for $t>0$ infinitesimally close to 0. From Eq.~\eqref{eq:Ckt_easy}, we find that 
\begin{multline}
\dot{C}_0 = \alpha \lim_{t_0\to\infty}\int_0^{t_0}dt'
\dot{\lambda}(t_0 -t')\lambda (t_0 -t')
= - 0.5\alpha \lim_{t_0\to\infty}\int_0^{t_0}dt' \frac{d}{dt'}
\lambda^2 (t_0 -t')\\
= - 0.5\alpha \lim_{t_0\to\infty}\left[\lambda^2 (0) - \lambda^2 (t_0)\right]
= - 0.5\alpha.
\label{eq:dC0}
\end{multline}
In order to evaluate $C_0$, we note that $\dot{C}_0$ also follows from Eq.~\eqref{eq:dCkt2} with $t=0$, yielding $\dot{C}_0 = - 0.5\alpha = - \tilde{\gamma}_{\tau} C(\tau) - \tilde{\gamma}_0 C_0$. Using $C(\tau) = C_0 \cosh\left(\Omega \tau\right) -0.5\alpha \Omega^{-1} \sinh\left(\Omega \tau\right)$ given  by Eq.~\eqref{eq:Cktini} for $t>0$, we finally get the desired result
\begin{equation}
C_0 = \sigma_v^2 = \frac{\alpha}{2}\frac{\Omega+\tilde{\gamma}_\tau \sinh(\Omega\tau)}{\Omega\left[\tilde{\gamma}_0 + \tilde{\gamma}_\tau \cosh(\Omega\tau)\right]}.
\label{eq:variancenofeed}
\end{equation}
For $\tilde{\gamma}_0 = 0$, we obtain $\Omega = i \sqrt{\tilde{\gamma}_\tau}$. Using the identities $\sinh(i x) = i \sin x$ and $\cosh(i x) = \cos x$, the formula~\eqref{eq:variancenofeed} can be written as
\begin{equation}
\sigma_v^2 = \frac{\alpha}{2\tilde{\gamma}_\tau}\frac{1+\sin(\tilde{\gamma}_\tau\tau)}{\cos(\tilde{\gamma}_\tau\tau)},
\label{eq:variancenofeedJP}
\end{equation}
which is the result derived in Refs.~\cite{Frank2003,Geiss2019}.


\end{document}